\documentclass[12pt]{article}
\usepackage{amsfonts,amssymb,amsmath,mathrsfs}
\usepackage{hyperref}
\newcommand{\beq}{\begin{equation}}
\newcommand{\eeq}{\end{equation}}
\newcommand{\vp}{\vphantom}
\newcommand{\hp}{\hphantom}
\newcommand{\wt}{\widetilde}

\newcommand{\pt}{\partial}
\newcommand{\bs}{\boldsymbol}
\newcommand{\al}{\alpha}
\newcommand{\bt}{\beta}
\newcommand{\g}{\gamma}
\newcommand{\G}{\Gamma}

\newcommand{\de}{\delta}

\newcommand{\la}{\lambda}
\newcommand{\La}{\Lambda}
\newcommand{\s}{\sigma}
\newcommand{\mc}{\mathcal}
\newcommand{\mr}{\mathrm}

\newcommand{\ads}{AdS_5\times S^5}
\newcommand{\coset}{PSU(2,2|4)/(SO(1,4)\times SO(5))}

\begin{document}
\begin{center}
\textbf{\Large Multitwistor mechanics of massless superparticle on $AdS_5\times S^5$ superbackground} \\[0.3cm]
{\large D.V.~Uvarov\footnote{E-mail: d\_uvarov@\,hotmail.com}}\\[0.2cm]
\textit{NSC Kharkov Institute of Physics and Technology,}\\ \textit{61108 Kharkov, Ukraine}\\[0.5cm]
\end{center}
\begin{abstract}
Supertwistors relevant to $\ads$ superbackground of IIB
supergravity are studied in the framework of the $D=10$ massless
superparticle model in the first-order formulation. Product
structure of the background suggests using $D=1+4$ Lorentz-harmonic
variables to express momentum components tangent to $AdS_5$ and
$D=5$ harmonics to express momentum components tangent to $S^5$
that yields eight-supertwistor formulation of the superparticle's
Lagrangian. We find incidence relations of the supertwistors with the $\ads$
superspace coordinates and the set of the quadratic constraints
that supertwistors satisfy. It is shown how using the constraints for the
(Lorentz-)harmonic variables it is possible to reduce
eight-supertwistor formulation to the four-supertwistor one.
Respective supertwistors agree with those introduced previously in
other models.  Advantage of the four-supertwistor formulation is
the presence only of the first-class constraints that facilitates
analysis of the superparticle model.
\end{abstract}
\setcounter{equation}{0}
\def\theequation{\thesection.\arabic{equation}}

\section{Introduction}

To study AdS/CFT correspondence \cite{Maldacena}, \cite{GKP98},
\cite{Witten:1998qj}, the best explored instance of which is
provided by the duality of the Type IIB superstring on $\ads$
superbackground and $N=4$ supersymmetric Yang-Mills theory on its
4-dimensional conformal boundary, it is commonly used the
(super)space setting. Both theories are invariant under the
$SU(2,2|4)$ symmetry supergroup that was one of the primary
arguments in favor of the duality. This symmetry supergroup is
manifest and linearly realized in supertwistor space
\cite{Penrose}, \cite{Ferber}. $D=4$ $N=4$ super-Yang-Mills theory
can also be described in supertwistor or superambitwistor space at
the level of supermultiplet \cite{Shirafuji}, field equations
\cite{Witten'78} and Lagrangian \cite{Witten'03}, \cite{Mason'05},
\cite{MS}, \cite{Boels}. Thus it is natural to wonder whether
$\ads$ superstring also admits a formulation in terms of
supertwistors, what are these supertwistors and what implications
such a formulation might have on the study of the $AdS_5/CFT_4$
correspondence. In the bosonic limit some of these issues were
addressed recently in \cite{Williams}.\footnote{Earlier discussion of the
role of supertwistors for $AdS_5/CFT_4$ correspondence can be found, e.g. in \cite{Siegel}.} In particular, there was given
the definition of twistors for $AdS_5$ space based on its
realization as a projective manifold. Another definition of
twistors \cite{Claus}, \cite{CKR}\footnote{See also subsequent
results in \cite{Cederwall}, \cite{ABGPKT16}, \cite{ABGPKT17}.}
follows from considering point particle mechanics on $AdS_5$ and
is based on the on-shell momentum realization as the product of
two constrained spinors. Generalization of these twistors to the
$\ads$ superbackground appears non-trivial since it is not
superconformally flat \cite{BILS}.

Independently I.~Bars in the framework of 2T approach \cite{BDA}
proposed superparticle and tensionless superstring models
\cite{Bars-twistor-string}, \cite{Bars-ads5s5} extended local
symmetries of which can be fixed in a variety of ways. One of the
gauges yields the first-order Lagrangian of the superparticle
(tensionless superstring) in $\ads$ superspace. In another gauge
dynamical variables of the model are supertwistors transforming in
the fundamental representation of the $SU(2,2|4)$ supergroup and
fundamental representation of its gauged subgroup. These models
provide an interesting alternative to the conventional way of
introducing (super)twistors in particle (string, brane) models
\cite{Ferber}, \cite{Shirafuji}, \cite{bbcl}-\cite{Bandos'14}.
However, it is unclear how the incidence relations that connect
supertwistor components and the superspace coordinates can be
derived within this approach.

This motivated us to examine in Ref.~\cite{U'18} the traditional root of introducing supertwistors starting with the first-order representation of the massless superparticle' Lagrangian on $\ads$ superbackground. Then we expressed momentum components tangent to $AdS_5$ and $S^5$ parts of the background in terms of the $SU(2)$-Majorana spinors for $Spin(1,4)$ and $Spin(5)$ groups so that the 10-momentum null condition was translated into the constraint on the sum of the products of these spinors. Substituting the definition of the supervielbein bosonic components as Cartan 1-forms of the $\coset$ supercoset allowed to transfer to the supertwistor form of the superparticle's Lagrangian. Respective $SU(2,2|4)$ supertwistors were assembled into $SU(2)$ doublets that differ from one another by the Grassmann parity of their components and coincide with those found in \cite{Bars-ads5s5}. The benefit of this approach is that it allows to find the incidence relations that connect the supertwistor components and $\ads$ supercoordinates via the $\coset$ representative. Explicit form of the incidence relations was derived for the particular example of the supercoset representative used to study the $\ads$ superstring mechanics \cite{MTlc}, \cite{MTT}, \cite{Metsaev'01}. Besides that, in the case when the superparticle's momentum components tangent to 5-sphere vanish, we quantized the model in terms of the superoscillators \cite{BG'83} and superambitwistors and showed how the supermultiplet of $D=5$ $N=8$ gauged supergravity \cite{GM'85} fits into its spectrum of states.

The geometric way of introducing spinors in the models of point-like and extended (supersymmetric) relativistic objects is based on the formulations of their Lagrangians that involve components of the local orthonormal repere \cite{VZ'85}. The repere matrix can be naturally identified with the element of the respective Lorentz group called vector Lorentz-harmonic matrix \cite{Sokatchev'86}. It can be expressed as the square of the spinor Lorentz-harmonic matrix \cite{DelducGS}, \cite{GalperinHS}, \cite{BZ'95} in analogy with the relation of the matrices of vector and spinor Lorentz transformations.
In the case of $D=1+3$ dimensions spinor Lorentz harmonics \cite{Bandos'90}, \cite{BZ'91} in the literature on twistors \cite{PR} were known as the normalized dyad.

The product structure of the bosonic $\ads$ background suggests using $D=1+4$ Lorentz harmonics \cite{Sokatchev'89}, \cite{U'15} to express momentum components in directions tangent to $AdS_5$ and $Spin(5)$ harmonics\footnote{Let us note that due to the local group isomorphism $Spin(5)\simeq USp(4)$, $Spin(5)$ harmonics that we use coincide with the $USp(4)$ harmonics introduced in \cite{Viet} and employed in \cite{Howe'94}, \cite{Ferrara-Sokatchev}, \cite{Zupnik'07}, \cite{JBIS}, \cite{Belyaev} to study harmonic superspace formulations for various supersymmetric field theories in $D=3,4,6$ dimensions.} for the momentum components in directions tangent to $S^5$. So one of the goals of the present paper is to derive eight-supertwistor formulation of the $D=10$ massless  superparticle on the $\ads$ superbackground starting with its first-order formulation that involves $D=1+4$ spinor Lorentz harmonics and $D=5$ harmonics. This explains the origin of the supertwistors that arise in the respective gauge for the 2T tensionless superstring model of Ref.~\cite{Bars-twistor-string} and allows to find their incidence relations with the $\ads$ superspace coordinates. We find the set of quadratic constraints that these supertwistors satisfy and calculate their classical algebra. It appears that some of the constraints are the second-class ones. This indicates the presence of the redundant d.o.f. and complicates further steps of the Dirac analysis of the model. Using the defining conditions for the spinor (Lorentz) harmonics we show that it is possible to exclude half of them from the expressions for the momentum components tangent to $AdS_5$ and $S^5$ parts of the background. Reduced expressions for the momentum components coincide with those proposed in \cite{U'18} and allow to obtain four-supertwistor formulation of the superparticle. Since these supertwistors precisely coincide with those found in \cite{Bars-ads5s5}, we actually find the relation between the eight- and four-supertwistor formulations of the massless superparticle on $\ads$ superbackground.

The organization of the paper is the following. In the next section as a warm up example we consider the model of the bosonic particle of mass $\mr m$ on $AdS_5$. We introduce vector and spinor Lorentz harmonics that parametrize the $SO(1,4)/SO(4)$ coset and express particle's momentum in terms of them. Then we transfer to the four-twistor formulation and find the set of the constraints that these twistors satisfy. It is shown that among them there are the second-class constraints. In section 3 using the harmonicity conditions we derive the reduced expression for the particle's momentum that depends just on the half of the spinor Lorentz harmonics. Corresponding twistors and two-twistor particle's Lagrangian are those of Ref.~\cite{Claus}. In \cite{CKR} this model, that is characterized only by the first-class constraints, was quantized in terms of the $SU(2)\times SU(2)$ bosonic oscillators and was shown to describe $SU(2,2)$ irreducible representation with spin zero and $AdS$ energy $E=2+|\mr m|$. We give alternative description of this representation in terms of the ambitwistors.

Section 4 is devoted to study of the massless superparticle on
$\ads$ superbackground. Momentum components tangent to $AdS_5$ and
$S^5$ are expressed in terms of the $Spin(1,4)$ Lorentz harmonics
and $Spin(5)$ harmonics. This allows to pass to the
eight-supertwistor formulation of the superparticle's
Lagrangian with the supertwistors being the elements of the
$(4|4)\times(4|4)$ supermatrix. We identify the set of the
constraints for supertwistors and calculate their classical
superalgebra to show that, similarly to the bosonic particle case,
there are present the second-class constraints. Then in section 5
it is shown how proposed in \cite{U'18} four-supertwistor formulation of the
superparticle's Lagrangian arises. We also consider the
set of equations for the quantized superparticle's wave function
in the superambitwistor space.

Necessary details of the spinor algebra and the algebra of the constraints for the eight-supertwistor formulation for the massless superparticle are collected in the appendices.

\setcounter{equation}{0}
\section{Massive bosonic particle on $AdS_5$}

As the starting point we choose the first-order form of the action of $D=5$ massive bosonic particle
\beq\label{d5-particle-action}
S_{AdS_5}=\int d\tau\mathscr L_{AdS_5},\quad \mathscr L_{AdS_5}=p_{m'}E^{m'}_\tau-\frac{g}{2}(p_{m'}p^{m'}+\mr m^2),
\eeq
where $E^{m'}_\tau$ is the world-line pull-back of the $AdS_5$ vielbein 1-form.
Particle's 5-momentum can be set proportional to the time-like unit-norm vector $n^{(0)}_{m'}$
\beq\label{ads-momentum-rep}
p_{m'}(\tau)=\mr mn^{(0)}_{m'}(\tau)
\eeq
that can be chosen as the first column of the $D=5$ vector Lorentz-harmonic matrix
\beq\label{vector-harmonics}
n^{(k')}_{m'}(\tau)=(n^{(0)}_{m'},n^{(\hat I)}_{m'})\in SO(1,4)\quad\Leftrightarrow\quad
n^{(k')}_{m'}\eta^{m'n'}n^{(l')}_{n'}=\eta^{(k')(l')}.
\eeq
It is identified with the local orthonormal repere in the tangent space to the particle's world line, the indices in brackets labelling the repere components. Orthonormality of the vector Lorentz harmonics implies that $n^{(0)}_{m'}n^{m'(0)}=-1$ so that the particle's mass-shell constraint is satisfied.
Vector Lorentz-harmonics transform under left global and right local Lorentz rotations
\beq
n^{(k')}_{m'}\quad\rightarrow\quad L_{m'}{}^{n'}n^{(l')}_{n'}R_{(l')}{}^{(k')},\quad L\in SO(1,4)_L,\quad R\in SO(1,4)_R.
\eeq
Because $n^{(0)}_{m'}$ is invariant only under the transformations from the subgroup $SO(4)_R\subset SO(1,4)_R$ that enters the set of the gauge symmetries of the massive particle's action, vector Lorentz harmonics parametrize the coset $SO(1,4)/SO(4)$.

To transfer to the twistor form of the particle's Lagrangian it is
necessary to introduce $Spin(1,4)$ Lorentz harmonics
\beq\label{spin14-harmonics}
v^{\bs\al}_{\bs\mu}=(-v^{\bs\al}_b,v^{\bs\al\dot b}) 
\eeq 
that can
be viewed as the 'square root' of the vector Lorentz harmonics
\beq n^{(k')}_{m'}=-\frac14v^{\mr T\bs\la}_{\hp{\mr
T}\bs\al}\g_{m'}{}^{\bs\al}{}_{\bs\bt}v^{\bs\bt}_{\bs\nu}\g^{(k')\bs\nu}{}_{\bs\la}.
\eeq 
We use letters from the beginning of the Greek alphabet to
label $Spin(1,4)_L$ spinor indices,
while those from the middle part of the Greek alphabet to label 
$Spin(1,4)_R$ spinor indices. Dotted and undotted letters from 
the beginning of the Latin alphabet label fundamental
representation indices of the two $SU(2)_R$ subgroups 
of $Spin(1,4)_R$. Details of the spinor
algebra in $D=1+4$ dimensions are given in \cite{U'18} and
Appendix A of the present paper.
In particular, for $n^{(0)}_{m'}$ we have
\beq\label{time-like-harmonic}
n^{(0)}_{m'}=-\frac14v^{\mr T\bs\la}_{\hp{\mr T}\bs\al}\g_{m'}{}^{\bs\al}{}_{\bs\bt}v^{\bs\bt}_{\bs\nu}\g^{(0)\bs\nu}{}_{\bs\la}=-\frac14(v_{\bs\al}^b\g_{m'}{}^{\bs\al}{}_{\bs\bt}v_b^{\bs\bt}-v_{\bs\al}^{\dot b}\g_{m'}{}^{\bs\al}{}_{\bs\bt}v_{\dot b}^{\bs\bt}).
\eeq
Spinor Lorentz harmonics satisfy the reality conditions
\beq\label{spin-lor-reality}
\g^{(0)\bs\la}{}_{\bs\nu}(v^{\bs\al}_{\bs\nu})^\dagger\g^{0\bs\al}{}_{\bs\bt}=v^{\mr T\bs\la}_{\hp{\mr T}\bs\bt}
\eeq
that ensure reality of the vector Lorentz harmonics. For the $4\times2$ blocks of the spinor Lorentz-harmonic matrix that carry fundamental representation indices of the two $SU(2)_R$ subgroups 
of $Spin(1,4)_R$ Eq.~(\ref{spin-lor-reality}) translates into the $SU(2)$-Majorana conditions
\beq\label{su2-major-cond}
(v^{\bs\al}_b)^\dagger\g^{0\bs\al}{}_{\bs\bt}=v^{\mr Tb}_{\hp{\mr T}\bs\bt},\quad(v^{\bs\al\dot b})^\dagger\g^{0\bs\al}{}_{\bs\bt}=v^{\mr T}_{\hp{\mr T}\bs\bt\dot b}.
\eeq
Orthonormality of the vector Lorentz harmonics then follows from the harmonicity conditions for the spinor Lorentz harmonics
\beq\label{harmonicity-cond}
v^{\bs\al}_{\bs\mu}v^{\bs\nu}_{\bs\al}=\de^{\bs\nu}_{\bs\mu}:\quad v^{\bs\al}_bv^c_{\bs\al}=-\de^c_b,\; v^{\bs\al\dot b}v_{\bs\al\dot c}=\de^{\dot b}_{\dot c},\; v^{\bs\al}_bv_{\bs\al\dot c}=0
\eeq
that state invariance of the charge conjugation matrices under the left and right $Spin(1,4)$ transformations since $v^{\bs\nu}_{\bs\al}=C_{\bs\al\bs\bt}C^{\bs\nu\bs\la}v^{\bs\bt}_{\bs\la}$ and ensure that $v^{\bs\al}_{\bs\mu}\in Spin(1,4)$.
Eq.~(\ref{time-like-harmonic}) is invariant under local $SU(2)_R\times SU(2)_R\subset Spin(1,4)_R$ symmetry so that spinor harmonics pertinent to the massive particle model parametrize the coset $Spin(1,4)/(SU(2)\times SU(2))$. This $SU(2)_R\times SU(2)_R$ gauge symmetry will also be the symmetry in the (super)twistor formulation.

Using the definition of $su(2,2)$ left-invariant Cartan forms
\beq\label{ads-cartan}
G^{-1\bs\al}{}_{\bs\g}dG^{\bs\g}{}_{\bs \bt}=\frac{i}{2}(E^{m'}(d)\g_{m'}{}^{\bs\al}{}_{\bs\bt}+E^{m'n'}(d)\g_{m'n'}{}^{\bs\al}{}_{\bs\bt})\in
su(2,2),
\eeq
where $G\in SU(2,2)$, 1-form, world-line pullback of which defines kinetic term of the particle's Lagrangian (\ref{d5-particle-action}), can be written in terms of $SU(2,2)$ twistors\footnote{In the first line spinor Lorentz harmonics were rescaled $v\to\sqrt{\frac{\mr m}{2}}v$ to bring the twistor kinetic term to the canonical form.}
\beq\label{bos-kin-term}
\begin{array}{rl}
p_{m'}E^{m'}(d)=&-\frac12v^{\mr T\bs\la}_{\hp{\mr T}\bs\g}\g_{m'}{}^{\bs\g}{}_{\bs\de}v^{\bs\de}_{\bs\nu}\g^{(0)\bs\nu}{}_{\bs\la}E^{m'}(d)\\[0.2cm]
=&iv^{\mr T\bs\la}_{\hp{\mr T}\bs\g}G^{-1\bs\g}{}_{\bs\al}dG^{\bs\al}{}_{\bs\de}v^{\bs\de}_{\bs\nu}\g^{(0)\bs\nu}{}_{\bs\la}\\[0.2cm]
=&\frac{i}{2}(\bar Z^{\bs\la}_{\bs\al}dZ^{\bs\al}_{\bs\nu}-d\bar Z^{\bs\la}_{\bs\al}Z^{\bs\al}_{\bs\nu})\g^{(0)\bs\nu}{}_{\bs\la}\\[0.2cm]
=&\frac{i}{2}(\bar Z^{a}_{\bs\al}dZ^{\bs\al}_{a}-d\bar Z^{a}_{\bs\al}Z^{\bs\al}_{a})+\frac{i}{2}(\bar Z_{\bs\al\dot a}dZ^{\bs\al\dot a}-d\bar Z_{\bs\al\dot a}Z^{\bs\al\dot a}).
\end{array}
\eeq
Incidence relations for the introduced twistors read
\beq\label{bos-incidence}
Z^{\bs\al}_{\bs\nu}=(-Z^{\bs\al}_b,\; Z^{\bs\al\dot b})=G^{\bs\al}{}_{\bs\bt}v^{\bs\bt}_{\bs\nu},\quad
\bar Z^{\bs\nu}_{\bs\al}=\left(
\begin{array}{c}
\bar Z^b_{\bs\al} \\
\bar Z_{\bs\al\dot b}
\end{array}
\right) =v^{\mr T\bs\nu}_{\hp{\mr T}\bs\bt}G^{-1\bs\bt}{}_{\bs\al}
\eeq 
and dual twistor matrix satisfies 
\beq\label{bos-hermit}
H^{\bs\la}{}_{\bs\nu}(Z^{\bs\al}_{\bs\nu})^\dagger
H^{\bs\al}{}_{\bs\bt}=\bar Z^{\bs\la}_{\bs\bt}. 
\eeq 
Matrices
$H^{\bs\al}{}_{\bs\bt}$ and $H^{\bs\la}{}_{\bs\nu}$ are the
metrics connecting fundamental and antifundamental
representations of $SU(2,2)_L$ and $SU(2,2)_R$ respectively. In
Ref.~\cite{U'18} there was chosen such realization of $D=2+4$
$\g$-matrices that $H^{\bs\al}{}_{\bs\bt}$ equals $D=1+4$ matrix
$\g^{0\bs\al}{}_{\bs\bt}$ and is off-diagonal
\beq\label{h-twistor} 
H^{\bs\al}{}_{\bs\bt}= \left(
\begin{array}{cc}
0 & \mr I_{2\times2} \\[0.2cm]
\mr I_{2\times2} & 0
\end{array}
\right)
\eeq
as is common in the twistor theory. For $\g^{(0)\bs\mu}{}_{\bs\nu}$ instead diagonal realization is used (see Appendix A) so that
\beq\label{h-diag}
H^{\bs\la}{}_{\bs\nu}=\g^{(0)\bs\la}{}_{\bs\nu}=\left(
\begin{array}{cc}
-\mr I_{2\times2} & 0 \\[0.2cm]
0 & \mr I_{2\times2}
\end{array}
\right).
\eeq
In this realization conjugation rules (\ref{bos-hermit}) for $4\times2$ blocks of the twistor matrix read 
\beq
(Z^{\bs\al}_b)^\dagger H^{\bs\al}{}_{\bs\bt}=\bar Z^b_{\bs\bt},\quad(Z^{\bs\al\dot b})^\dagger H^{\bs\al}{}_{\bs\bt}=\bar Z_{\bs\bt\dot b}.
\eeq
Using the incidence relations (\ref{bos-incidence}) and the harmonicity conditions (\ref{harmonicity-cond}) one derives the constraints that twistors obey
\beq\label{tw-constr}
\bar Z^{\bs\la}_{\bs\al}Z^{\bs\al}_{\bs\nu}=v^{\mr T\bs\la}_{\hp{\mr T}\bs\al}v^{\bs\al}_{\bs\nu}=\frac{\mr m}{2}\de^{\bs\la}_{\bs\nu}
\eeq
so that $\sqrt{\frac{2}{\mr m}}Z^{\bs\al}_{\bs\nu}\in U(2,2)$ modulo the gauge symmetries of the particle's action in twistor formulation
\beq\label{bos-twistor-action}
S_{AdS_5}=\int d\tau\mathscr L_{4-\mr{twistor}},\quad \mathscr L_{4-\mr{twistor}}=\frac{i}{2}(\bar Z^{\bs\la}_{\bs\al}\dot Z^{\bs\al}_{\bs\nu}-\dot{\bar Z}^{\bs\la}_{\bs\al}Z^{\bs\al}_{\bs\nu})\g^{(0)\bs\nu}{}_{\bs\la}+\La^{\bs\nu}{}_{\bs\la}\left(\bar Z^{\bs\la}_{\bs\al}Z^{\bs\al}_{\bs\nu}-\frac{\mr m}{2}\de^{\bs\la}_{\bs\nu}\right).
\eeq
The structure of the kinetic term of this action is similar to that of the tensionless superstring of Ref.~\cite{Bars-twistor-string} in the twistor gauge.  It has manifest global $SU(2,2)_L$ symmetry, whereas $SU(2,2)_R$ symmetry, assumed in the definition of the dual twistor matrix (\ref{bos-hermit}), is broken by $\g^{(0)\bs\nu}{}_{\bs\la}$.

It is possible to give another definition of the dual twistor matrix
\beq\label{other-dual-twistor}
\wt Z^{\bs\la}_{\bs\al}=\g^{(0)\bs\la}{}_{\bs\nu}\bar Z^{\bs\nu}_{\bs\al}=\left(
\begin{array}{c}
-\bar Z^b_{\bs\al} \\
\bar Z_{\bs\al\dot b}
\end{array}
\right)=(Z^{\bs\bt}_{\bs\la})^\dagger H^{\bs\bt}{}_{\bs\al}
\eeq
that differs from (\ref{bos-hermit}) by the multiplication by $\g^{(0)\bs\nu}{}_{\bs\la}$. Since $\g^{(0)}\g^{(0)}=1$,
the kinetic term of the twistor Lagrangian (\ref{bos-twistor-action}) acquires the following form in terms of the dual twistor matrix (\ref{other-dual-twistor})
\beq
\frac{i}{2}(\wt Z^{\bs\la}_{\bs\al}\dot Z^{\bs\al}_{\bs\la}-\dot{\wt Z}{}^{\bs\la}_{\bs\al}Z^{\bs\al}_{\bs\la})=\frac{i}{2}(\bar Z^{a}_{\bs\al}\dot Z^{\bs\al}_{a}-\dot{\bar Z}^{a}_{\bs\al}Z^{\bs\al}_{a})+\frac{i}{2}(\bar Z_{\bs\al\dot a}\dot Z^{\bs\al\dot a}-\dot{\bar Z}_{\bs\al\dot a}Z^{\bs\al\dot a}).
\eeq
It has manifest $SU(2)_R\times SU(2)_R$ gauge symmetry but actually is invariant under the $SU(4)_R$ symmetry
that is broken by the constraint term
\beq
\La^{\bs\nu}{}_{\bs\la}\left(\wt Z^{\bs\la}_{\bs\al}Z^{\bs\al}_{\bs\nu}-\frac{\mr m}{2}\g^{(0)\bs\la}{}_{\bs\nu}\right).
\eeq

To analyze the algebra of the constraints let us decompose them into the $SU(2)_R\times SU(2)_R$ irreducible constituents
\beq
\begin{array}{c}
L^a{}_b=\bar Z^a_{\bs\al}Z^{\bs\al}_b-\frac12\de^a_b\bar Z^c_{\bs\al}Z^{\bs\al}_c\approx0,\quad
M_{\dot a}{}^{\dot b}=\bar Z_{\bs\al\dot a}Z^{\bs\al\dot b}-\frac12\de_{\dot a}^{\dot b}\bar Z_{\bs\al\dot c}Z^{\bs\al\dot c}\approx0,\\[0.2cm]
\mr e=\bar Z^a_{\bs\al}Z^{\bs\al}_a+\bar Z_{\bs\al\dot a}Z^{\bs\al\dot a}\approx0,\quad
\mr c=\bar Z^a_{\bs\al}Z^{\bs\al}_a-\bar Z_{\bs\al\dot a}Z^{\bs\al\dot a}+2\mr m\approx0,\\[0.2cm]
\bar Z^a_{\bs\al}Z^{\bs\al\dot b}\approx0,\quad\bar Z_{\bs\al\dot a}Z^{\bs\al}_b\approx0.
\end{array}
\eeq
Basic Dirac bracket (D.B.) relations for the twistor components that follow from (\ref{bos-twistor-action}) are
\beq\label{dir-br}
\{Z^{\bs\al}_a,\bar Z^b_{\bs\bt}\}_{D.B.}=i\de^b_a\de^{\bs\al}_{\bs\bt},\quad
\{Z^{\bs\al\dot a},\bar Z_{\bs\bt\dot b}\}_{D.B.}=i\de_{\dot b}^{\dot a}\de^{\bs\al}_{\bs\bt}.
\eeq
Then non-trivial D.B. relations of the constraints acquire the form 
\beq
\begin{array}{c}
\{\bar Z^a_{\bs\al}Z^{\bs\al\dot b},\bar Z_{\bs\bt\,\dot a}Z^{\bs\bt}_b\}_{D.B.}=i\de^{\dot b}_{\dot a}L^a{}_b-i\de^a_bM_{\dot a}{}^{\dot b}+\frac{i}{2}\de^a_b\de^{\dot b}_{\dot a}(\mr c-2\mr m),\\[0.2cm]
\{L^a{}_b,\bar Z_{\bs\al\dot a}Z^{\bs\al}_c\}_{D.B.}\!=\!\frac{i}{2}\de^a_b\bar Z_{\bs\al\dot a}Z^{\bs\al}_c-i\de^a_c\bar Z_{\bs\al\dot a}Z^{\bs\al}_b,\quad
\{L^a{}_b,\bar Z_{\bs\al}^cZ^{\bs\al\dot b}\}_{D.B.}\!=\!-\frac{i}{2}\de^a_b\bar Z_{\bs\al}^cZ^{\bs\al\dot b}+i\de^c_b\bar Z_{\bs\al}^aZ^{\bs\al\dot b},\\[0.2cm]
\{M_{\dot a}{}^{\dot b},\bar Z_{\bs\al\dot c}Z^{\bs\al}_b\}_{D.B.}\!=\!-\frac{i}{2}\de_{\dot a}^{\dot b}\bar Z_{\bs\al\dot c}Z^{\bs\al}_b+i\de_{\dot c}^{\dot b}\bar Z_{\bs\al\dot a} Z^{\bs\al}_b,\quad
\{M_{\dot a}{}^{\dot b},\bar Z_{\bs\al}^aZ^{\bs\al\dot c}\}_{D.B.}\!=\!\frac{i}{2}\de_{\dot a}^{\dot b}\bar Z_{\bs\al}^aZ^{\bs\al\dot c}-i\de^{\dot c}_{\dot a}\bar Z_{\bs\al}^aZ^{\bs\al\dot b},\\[0.2cm]
\{L^a{}_b,L^c{}_d\}_{D.B.}=i(\de^c_bL^a{}_d-\de^a_dL^c{}_b),\quad
\{M_{\dot a}{}^{\dot b},M_{\dot c}{}^{\dot d}\}_{D.B.}=i(\de^{\dot b}_{\dot c}M_{\dot a}{}^{\dot d}-\de_{\dot a}^{\dot d}M_{\dot c}{}^{\dot b}),\\[0.2cm]
\{\mr c,\bar Z^a_{\bs\al}Z^{\bs\al\dot b}\}_{D.B.}=2i\bar Z^a_{\bs\al}Z^{\bs\al\dot b},\quad
\{\mr c,\bar Z_{\bs\al\dot a}Z^{\bs\al}_b\}_{D.B.}=-2i\bar Z_{\bs\al\dot a}Z^{\bs\al}_b.
\end{array}
\eeq
These are relations of the $SU(4)_R$ symmetry algebra broken by $\mr m$. Thus the constraints $\bar Z^a_{\bs\al}Z^{\bs\al\dot b}$ and $\bar Z_{\bs\al\dot a}Z^{\bs\al}_b$ are the second-class, while other eight constraints are the first-class ones. From the viewpoint of the oscillator realization of the $SU(4)_R$ algebra $\bar Z^a_{\bs\al}Z^{\bs\al\dot b}$ and $\bar Z_{\bs\bt\dot a}Z^{\bs\bt}_b$ are raising and lowering generators from the $+1$ and $-1$ eigenspaces w.r.t. to the three-grading structure introduced by $\mr c$ \cite{GMZ2}. Having identified the first- and second-class constraints it is possible to verify that the number of the physical degrees of freedom in the four-twistor formulation equals eight matching that of the space-time formulation of the $D=5$ massive particle mechanics.

\setcounter{equation}{0}
\section{Reduction of four-twistor formulation to two-twistor formulation}

The fact that in the four-twistor formulation there are eight
second-class constraints quadratic in twistors essentially
complicates further analysis of the Hamiltonian mechanics of the
bosonic particle model and requires introduction of the D.B. for
these constraints. It is, however, possible to remove some of the
redundant degrees of freedom that will simplify the algebra of the
remaining constraints without breaking the $SU(2,2)_L$ invariance.
The form of the incidence relations (\ref{bos-incidence}) shows
that reduction of the $SO(1,4)$ Lorentz-harmonic variables
underlies reduction of twistors, hence we discuss it on the level
of harmonics. To this end decompose spinor Lorentz harmonics into
$SL(2,\mathbb C)_L$ constituents 
\beq v^c_{\bs\al}=\left(
\begin{array}{c}
v^c_{\al} \\ \bar v^{\dot\al c}
\end{array}
\right),
\quad
v^{\dot c}_{\bs\al}=\left(
\begin{array}{c}
v^{\dot c}_{\al} \\ \bar v^{\dot\al\dot c}
\end{array}
\right).
\eeq
$SU(2)$-Majorana conditions that these harmonics satisfy
\beq
(v^c_{\bs\al})^\dagger\g^{0}{}_{\bs\al}{}^{\bs\bt}=-v^{\mr T\bs\bt}_{\hp{\mr T}c},\quad(v^{\dot c}_{\bs\al})^\dagger\g^{0}{}_{\bs\al}{}^{\bs\bt}=v^{\mr T\bs\bt}_{\hp{\mr T}\dot c},
\eeq
when expressed in terms of the $SL(2,\mathbb C)_L$ constituents read
\beq
(v^c_\al)^\dagger=\bar v_{\dot\al c},\quad(v^{\dot c}_\al)^\dagger=-\bar v_{\dot\al\dot c}.
\eeq
Let us define
\beq
v^\al_c v^c_{\al}=\Upsilon\in\mathbb C\setminus\{0\},\quad v^\al_{\dot c} v^{\dot c}_{\al}=\tilde\Upsilon\in\mathbb C\setminus\{0\}.
\eeq
Then from the harmonicity conditions we obtain
\beq\label{det-mass}
\Upsilon+\bar\Upsilon=\tilde\Upsilon+\bar{\tilde\Upsilon}=\mr m
\eeq
and
\beq\label{4-harm}
v^\al_av_{\al\dot c}+\bar v^{\dot\al}_a\bar v_{\dot\al\dot c}=0.
\eeq
Multiplying the latter relation by $v^a_\bt$ we derive
\beq\label{dot-vs-undot}
v_{\bt\dot c}=\frac{2}{\Upsilon}v^a_\bt\bar v_{\dot\al a}\bar v^{\dot\al}_{\dot c}.
\eeq
Further multiplying by $v^{\bt\dot c}$ gives $\Upsilon\tilde\Upsilon=\bar\Upsilon\bar{\tilde\Upsilon}$ and in view of (\ref{det-mass})
\beq\label{ups}
\Upsilon=\bar{\tilde\Upsilon}.
\eeq
Then using (\ref{dot-vs-undot}) and (\ref{ups}) it is possible to show that in the expression for the particle's 5-momentum 
\beq
p_{m'}=-\frac12v_{\bs\al}^b\g_{m'}{}^{\bs\al}{}_{\bs\bt}v_b^{\bs\bt}+\frac12v_{\bs\al}^{\dot b}\g_{m'}{}^{\bs\al}{}_{\bs\bt}v_{\dot b}^{\bs\bt}=\left\{
\begin{aligned}
p_m&=v^{\al b}\sigma_{m\al\dot\al}\bar v^{\dot\al}_b+v^{\al}_{\dot b}\sigma_{m\al\dot\al}\bar v^{\dot\al\dot b} \\
p_5&=\Upsilon_I-\tilde\Upsilon_I
\end{aligned}\right.
\eeq
contributions of the first and the second summands equal so that one can exclude spinor Lorentz harmonics with dotted indices
\beq\label{reduced-5-momentum}
p_{m'}=-\frac12v_{\bs\al}^b\g_{m'}{}^{\bs\al}{}_{\bs\bt}v_b^{\bs\bt},
\eeq
where the factor of two has been absorbed in the definition of Lorentz harmonics with undotted indices that now satisfy
\beq
v_b^{\bs\al}v^b_{\bs\al}=-2m.
\eeq
Let us remark that using (\ref{det-mass}) and (\ref{4-harm}) it is possible to exclude in a similar way spinor Lorentz harmonics with undotted indices leaving only those with dotted indices. Also note that in the expressions for the space-like vector-columns of the Lorentz-harmonic matrix (\ref{vector-harmonics}) it is impossible to remove contributions of the spinor harmonics with either dotted or undotted indices.

Taking expression for the reduced 5-momentum (\ref{reduced-5-momentum}) one can obtain two-twistor form of the massive particle's Lagrangian
following the same steps as in (\ref{bos-kin-term})
\beq\label{2-twistor-lagr}
\mathscr L_{2-\mr{twistor}}=\frac{i}{2}(\bar Z^a_{\bs\al}\dot Z^{\bs\al}_a-\dot{\bar Z}{}^a_{\bs\al}Z^{\bs\al}_a)+\La^a{}_bL^b{}_a+\La(\bar Z^a_{\bs\al}Z^{\bs\al}_a+2\mr m)
\eeq
that was proposed in \cite{Claus} (see also \cite{ABGPKT16}, \cite{ABGPKT17}). In \cite{CKR}, using the isomorphism of the algebras of quantized twistors and $su(2)\oplus su(2)$ bosonic oscillators that are employed in constructing positive energy unitary irreps of $SU(2,2)$ and its superextensions \cite{BG'83}, \cite{GM'85}, it was shown that this model describes $SU(2,2)$ unitary irrep with $AdS$ energy $E=2+|\mr m|$ and zero spin. One needs at least two copies of $su(2)\oplus su(2)$ bosonic oscillators to describe such irreps. Their lowest-weight vectors are given by the antisymmetrized products of creation oscillators in the fundamental representation of one of the $su(2)$ algebras \cite{GMZ2}, \cite{GMZ1}.

In this section we construct particle's wave function in the ambitwistor space 
corresponding to such irrep. Let us
introduce specific notation for each of the twistors 
\beq
Z^{\bs\al}_1\equiv Z^{\bs\al},\quad Z^{\bs\al}_2\equiv
W^{\bs\al},\quad\bar Z^1_{\bs\al}\equiv\bar Z_{\bs\al},\quad\bar
Z^2_{\bs\al}\equiv\bar W_{\bs\al}. 
\eeq 
It is readily shown that
from (\ref{2-twistor-lagr}) there follow conventional commutation
relations of the quantized twistors 
\beq 
[Z^{\bs\al},\bar
Z_{\bs\bt}]=\de^{\bs\al}_{\bs\bt},\quad [W^{\bs\al},\bar
W_{\bs\bt}]=\de^{\bs\al}_{\bs\bt}. 
\eeq 
Hermitian operators
corresponding to four first-class constraints of the model can be
brought to the form 
\beq\label{ambitwistor-constr} 
Z\bar
Z+m-2\approx0,\quad\bar WW+m+2\approx0,\quad\bar
ZW\approx0,\quad\bar WZ\approx0. 
\eeq 
Let $Z^{\bs\al}$ and $\bar
W_{\bs\al}$ be coordinates of the ambitwistor space, then other
two quantized twistors are represented as the first-order
differential operators 
\beq 
\bar Z_{\bs\al}=-\frac{\pt}{\pt
Z^{\bs\al}},\quad W^{\bs\al}=\frac{\pt}{\pt\vp{\hat
W_{\bs\al}}\bar W_{\bs\al}}. 
\eeq 
Imposing a la Gupta-Bleuler
three of the four constraints (\ref{ambitwistor-constr}) that are
given by the utmost first-order differential operators on the
particle's wave function $F_{(\mr m-2|-\mr m-2)}(Z,\bar W)$, we
find 
\beq
\begin{array}{c}
Z\frac{\pt}{\pt Z}F_{(\mr m-2|-\mr m-2)}(Z,\bar W)=(\mr m-2)F_{(\mr m-2|-\mr m-2)}(Z,\bar W),\\[0.2cm]
\bar W\frac{\pt}{\pt\vp{\hat{\bar W}}\bar W}F_{(\mr m-2|-\mr m-2)}(Z,\bar W)=-(\mr m+2)F_{(\mr m-2|-\mr m-2)}(Z,\bar W),\\[0.2cm]
(\bar WZ)F_{(\mr m-2|-\mr m-2)}(Z,\bar W)=0.
\end{array}
\eeq
These equations imply that particle's wave function is homogeneous in each of its arguments as indicated by the subscript and that its mass should be integer $\mr m\in\mathbb Z$ (cf. \cite{CKR}). The last constraint can be taken into account by writing
\beq
F_{(\mr m-2|-\mr m-2)}(Z,\bar W)=\de(\bar WZ)f_{(\mr m-1|-\mr m-1)}(Z,\bar W).
\eeq
When $\mr m$ is zero, it is known that the ambitwistor transform \cite{Eastwood-AMS}, \cite{Mason'13} of $f_{(-1|-1)}(Z,\bar W)$ as a function of Penrose twistors, corresponding to the boundary limit \cite{U'18} of $AdS_5$ twistors, gives off-shell scalar field on $D=4$ Minkowski space-time. This field can be identified as the $D=4$ shadow field \cite{Metsaev08} that serves as the boundary value of the non-normalizable solution of the Dirichlet problem for  $AdS_5$ scalar field. As discussed in \cite{U'16}, \cite{U'18} associated $SU(2,2)$ lowest-weight vector is given by the oscillator vacuum $|0\rangle$. If $\mr m>0$,
then the respective $SU(2,2)$ lowest-weight vector is
\beq
a^{[\al_1}(1)a^{\al_2]}(2)\cdots a^{[\al_{2\mr m-1}}(1)a^{\al_{2\mr m}]}(2)|0\rangle,
\eeq
as was found in \cite{CKR}. In ambitwistor description each of the antisymmetrized products of $a$-oscillators maps to the first-order differential operator \cite{U'18}
\beq
Z^{\bs\al}I_{\bs\al\bs\bt}\frac{\pt}{\pt\vp{\hat W_{\bs\bt}}\bar W_{\bs\bt}},
\eeq
so that massive particle's wave function in ambitwistor space is
\beq
f_{(\mr m-1|-\mr m-1)}(Z,\bar W)=\left(ZI\frac{\pt}{\pt\vp{\hat{\bar W}}\bar W}\right)^{\mr m}f_{(-1|-1)}(Z,\bar W),\quad\mr m\geq0.
\eeq
Respectively for $\mr m<0$ associated $SU(2,2)$ lowest-weight vector is
\beq
b_{[\al_1}(1)b_{\al_2]}(2)\cdots b_{[\al_{-2\mr m-1}}(1)b_{\al_{-2\mr m}]}(2)|0\rangle.
\eeq
Antisymmetrized products of $b$-oscillators map to the operator
\beq
\bar W_{\bs\al}I^{\bs\al\bs\bt}\frac{\pt}{\pt Z^{\bs\bt}}
\eeq
and particle's wave function equals
\beq
f_{(\mr m-1|-\mr m-1)}(Z,\bar W)=\left(\bar WI\frac{\pt}{\pt Z}\right)^{-\mr m}f_{(-1|-1)}(Z,\bar W),\quad\mr m\leq0.
\eeq
Note that differential operators $\left(ZI\frac{\pt}{\pt\vp{\hat{\bar W}}\bar W}\right)$ and $\left(\bar WI\frac{\pt}{\pt Z}\right)$ are used in twistor theory to describe massive fields in four-dimensional Minkowski space \cite{MCP}.

\setcounter{equation}{0}
\section{Eight-supertwistor formulation of massless superparticle on $\ads$ superbackground}

To derive eight-supertwistor formulation of the $D=10$ massless 
superparticle on the $AdS_5\times S^5$ superbackground let us
introduce diagonal supermatrix 
\beq\label{spinor-smatrix} \mr
V^{\mc A}{}_{\mc N}=\left(
\begin{array}{cc}
v^{\bs\al}_{\bs\nu} & 0\\[0.2cm]
0 & \ell^A_{N}
\end{array}
\right)
=\left(
\begin{array}{cccc}
-v^{\bs\al}_{b} & v^{\bs\al\dot b} & 0 & 0\\[0.2cm]
0 & 0 & \ell^A_{p} & \ell^{A\dot p}
\end{array}
\right).
\eeq
Its upper-diagonal block constitutes $4\times4$ matrix with $Spin(1,4)$ indices 
\beq\label{spin14-variables}
v^{\bs\al}_{\bs\mu}=(-v^{\bs\al}_b,v^{\bs\al\dot b}),
\eeq
while in the lower-diagonal block there is $4\times4$ matrix with $Spin(5)\sim USp(4)$ indices\footnote{We slightly abuse the notation in this section since, as will be seen below, $v^{\bs\al}_{\bs\mu}$ contains some additional degrees of freedom compared to the $Spin(1,4)$ Lorentz harmonics and $\ell^A_{N}$ has more independent degrees of freedom than the $USp(4)$ harmonics \cite{Viet}.}
\beq\label{spin5-variables}
\ell^A_{N}=(\ell^A_{p},\;\ell^{A\dot p}).
\eeq
Reality conditions for $Spin(1,4)$ variables coincide with those in (\ref{spin-lor-reality}) and for $Spin(5)$ variables are 
\beq
(\ell^A_L)^\dagger=\ell^{\mr T\,L}_{\hp{\mr T}A}:\quad(\ell^A_q)^\dagger=\ell^{\mr T\,q}_{\hp{\mr T}A},\quad(\ell^{A\dot q})^\dagger=\ell^{\mr T}_{A\dot q}.
\eeq
In the supermatrix form we have 
\beq
\mc H^{\mc L}{}_{\mc N}(\mr V^{\mc A}{}_{\mc N})^\dagger\mc H^{\mc A}{}_{\mc B}=\mr V^{\mr T\mc L}{}_{\mc B}
=\left(
\begin{array}{cc}
v^{\mr T\bs\la}_{\hp{\mr T}\bs\bt} & 0 \\[0.2cm]
0 & \ell^{\mr TL}_{\hp{\mr T}B}
\end{array}
\right),
\eeq
where
\beq
\mc H^{\mc A}{}_{\mc B}=\left(
\begin{array}{cc}
H^{\bs\al}{}_{\bs\bt} & 0 \\[0.2cm]
0 & \de^A_B
\end{array}
\right),\quad
\mc H^{\mc L}{}_{\mc N}=\left(
\begin{array}{cc}
H^{\bs\la}{}_{\bs\nu} & 0 \\[0.2cm]
0 & \de^L_N
\end{array}
\right)
\eeq
with $H^{\bs\al}{}_{\bs\bt}$ and $H^{\bs\la}{}_{\bs\nu}$ given (\ref{h-twistor}) and (\ref{h-diag}).
Introduce also the $\coset$ representative
\beq
\mc G^{\mc A}{}_{\mc B}=\left(
\begin{array}{cc}
G^{\bs\al}{}_{\bs\bt} & G^{\bs\al}{}_{B} \\[0.2cm]
G^A{}_{\bs\bt} & G^{A}{}_{B}
\end{array}
\right)\in\coset
\eeq
that satisfies
\beq
(\mc G^{\mc A}{}_{\mc B})^\dagger=\mc H^{\mc B}{}_{\mc C}\mc G^{-1\mc C}{}_{\mc D}\mc H^{\mc D}{}_{\mc A}.
\eeq
Define twistor supermatrix 
\beq\label{def-stwistor}
\mc Z^{\mc A}{}_{\mc N}=G^{\mc A}{}_{\mc B}\mr V^{\mc B}{}_{\mc N}=\left(\mc Z^{\mc A}_{\bs\nu}\;\Psi^{\mc A}_{N}\right)=\left(
-\mc Z^{\mc A}_{b}\;\mc Z^{\mc A\dot b}\;\Psi^{\mc A}_{p}\;\Psi^{\mc A\dot p}\right)
=\left(
\begin{array}{cccc}
-Z^{\bs\al}_b & Z^{\bs\al\dot b} & \xi^{\bs\al}_p & \xi^{\bs\al\dot p} \\
-\eta^A_b & \eta^{A\dot b} & L^A_p & L^{A\dot p}
\end{array}
\right)
\eeq
and its dual by 
\beq\label{def-dual-stwistor}
\bar{\mc Z}^{\mc L}{}_{\mc B}=\mc H^{\mc L}{}_{\mc N}(\mc Z^{\mc A}{}_{\mc N})^\dagger\mc H^{\mc A}{}_{\mc B}=\mr V^{\mr T\mc L}{}_{\mc C}\mc G^{-1\mc C}{}_{\mc B}=\left(
\begin{array}{c}
\bar{\mc Z}^{\bs\la}_{\mc B}\\
\bar\Psi^L_{\mc B}
\end{array}
\right)=\left(
\begin{array}{c}
\bar Z^b_{\mc B} \\
\bar Z_{\mc B\dot b} \\
\bar\Psi^q_{\mc B} \\
\bar\Psi_{\mc B\dot q}
\end{array}
\right)
=\left(
\begin{array}{cc}
\bar Z^b_{\bs\bt} & \bar\eta^b_B \\[0.2cm]
\bar Z_{\bs\bt\dot b} & \bar\eta_{B\dot b} \\[0.2cm]
\bar\xi^q_{\bs\bt} & \bar L^q_B \\[0.2cm]
\bar\xi_{\bs\bt\dot q} & \bar L_{B\dot q}
\end{array}
\right). 
\eeq 
For individual rows of the dual twistor supermatrix
this definition translates into the relations 
\beq 
\bar{\mc
Z}^b_{\mc B}=(\mc Z^{\mc A}_b)^\dagger\mc H^{\mc A}{}_{\mc
B},\quad\bar{\mc Z}_{\mc B\dot b}=(\mc Z^{\mc A\dot b})^\dagger\mc
H^{\mc A}{}_{\mc B},\quad \bar\Psi^q_{\mc B}=(\Psi^{\mc
A}_q)^\dagger\mc H^{\mc A}{}_{\mc B},\quad \bar\Psi_{\mc B\dot
q}=(\Psi^{\mc A\dot q})^\dagger\mc H^{\mc A}{}_{\mc B}. 
\eeq
Supertwistors $\mc Z^{\mc A}_b$, $\mc Z^{\mc A\dot b}$ and their
duals are conventional ones and were named $c$-type in
\cite{U'18}. Their $SU(2,2)_L$ components $Z^{\bs\al}_b$,
$Z^{\bs\al\dot b}$ are even and $SU(4)_L$ components $\eta^A_b$,
$\eta^{A\dot b}$ are odd. On the contrary supertwistors $\Psi^{\mc
A}_{p}$, $\Psi^{\mc A\dot p}$ and their duals are $a$-type since
their $SU(2,2)_L$ components $\xi^{\bs\al}_p$, $\xi^{\bs\al\dot
p}$ are odd but $SU(4)_L$ components $L^A_p$, $L^{A\dot p}$ are
even. Appearance of such unconventional supertwistors for
superparticle model on $AdS_5\times S^5$ superbackground was
discussed in \cite{Bars-twistor-string} and \cite{Bars-ads5s5}.
Note that the definition of the twistor supermatrix
(\ref{def-stwistor}) assumes that it transforms under left and
right $SU(2,2|4)$ transformations 
\beq 
\mc Z^{\mc A}{}_{\mc
N}\quad\to\quad\mc L^{\mc A}{}_{\mc B}\mc Z^{\mc B}{}_{\mc M}\mc
R^{\mc M}{}_{\mc N}, 
\eeq 
where supermatrices $\mc L^{\mc
A}{}_{\mc B}$ and $\mc R^{\mc M}{}_{\mc N}$ have $SU(2,2)$
upper-diagonal blocks and $SU(4)$ lower-diagonal blocks.

Consider now the first-order form of the massless superparticle
action on the $AdS_5\times S^5$ superbackground
\beq\label{d10sparticle} 
S_{AdS_5\times S^5}=\int d\tau\mathscr
L_{AdS_5\times S^5},\quad\mathscr L_{AdS_5\times
S^5}=p_{m'}E^{m'}_\tau+p_{I'}E^{I'}_\tau-\frac{g}{2}(p_{m'}p^{m'}+p_{I'}p^{I'}).
\eeq 
$E^{m'}_\tau$ and $E^{I'}_\tau$ are the world-line pullbacks
of the $D=10$ supervielbein bosonic components tangent to $AdS_5$
and $S^5$. In analogy with the discussion of section 2, momentum
components tangent to $AdS_5$ can be expressed as
\beq\label{ads-momentum} 
p_{m'}=-\frac12 v^{\mr T\bs\la}_{\hp{\mr
T}\bs\al}\g_{m'}{}^{\bs\al}{}_{\bs\bt}v^{\bs\bt}_{\bs\nu}\g^{(0)\bs\nu}{}_{\bs\la}=-\frac12(v_{\bs\al}^b\g_{m'}{}^{\bs\al}{}_{\bs\bt}v_b^{\bs\bt}-v_{\bs\al}^{\dot
b}\g_{m'}{}^{\bs\al}{}_{\bs\bt}v_{\dot b}^{\bs\bt}), 
\eeq 
where
$Spin(1,4)$ spinors $v^{\bs\al}_b$ and $v^{\bs\al\dot b}$
satisfy $SU(2)$-Majorana conditions (\ref{su2-major-cond}) and
also 
\beq\label{spin14-constr} v^{\bs\al}_bv_{\bs\al}^{\dot b}=0
\eeq 
so that 
\beq\label{ads-momentum-squared}
p_{m'}p^{m'}=-\frac14(v^{\bs\al}_av^a_{\bs\al}+v^{\bs\al}_{\dot
a}v^{\dot a}_{\bs\al})^2. 
\eeq 
Similarly momentum components
tangent to $S^5$ can be expressed 
\beq\label{s5-momentum}
p_{I'}=\frac12\ell^{\mr TL}_{\hp{\mr
T}A}\g_{I'}{}^A{}_B\ell^B_N\g^{(5)N}{}_L=-\frac12(\ell_A^{\, q}\g_{I'}{}^{A}{}_B\ell_q^B+\ell_A^{\,\dot
q}\g_{I'}{}^{A}{}_B\ell_{\dot q}^B) 
\eeq 
in terms of the $Spin(5)$
spinor variables (\ref{spin5-variables}) constrained by the
relations 
\beq\label{spin5-constr} 
\ell^A_q\ell^{\,\dot q}_A=0 
\eeq
necessary to obtain the following expression for the square of
momentum components tangent to $S^5$
\beq\label{sphere-momentum-squared}
p_{I'}p_{I'}=\frac14(\ell^A_q\ell^{\, q}_A-\ell^A_{\dot q}\ell^{\,\dot
q}_A)^2. 
\eeq 
As a result the null-momentum condition in 10
dimensions translates into 
\beq\label{10d-mass-shell-constr}
(v^{\bs\al}_av^a_{\bs\al}+v^{\bs\al}_{\dot a}v^{\dot
a}_{\bs\al})^2=(\ell^A_q\ell^{\, q}_A-\ell^A_{\dot q}\ell^{\,\dot
q}_A)^2. 
\eeq 
Let us note that null 10-momentum $(p_{m'},p_{I'})$
has 9 independent components, whereas at this point spinors
$(v^{\bs\al}_a,v^{\bs\al}_{\dot a})$ and $(\ell^A_q,\ell^A_{\dot
q})$ have $6+6-1=11$ independent components. Two additional
constraints for the spinor variables will be set below. Now 1-form
that enters kinetic term of the superparticle's Lagrangian
(\ref{d10sparticle}) can be expressed in the supertwistor form
\beq\label{kin-term-ph-vs-stwistor}
\begin{array}{rl}
p_{m'}E^{m'}(d)+p_{I'}E^{I'}(d)=&iV^{T\mc L}{}_{\mc D}\mc G^{-1}{}^{\mc D}{}_{\mc A}d\mc G^{\mc A}{}_{\mc C}V^{\mc C}{}_{\mc N}\G^{\mc N}{}_{\mc L}\\[0.2cm]
=&\frac{i}{2}\left(\bar{\mc Z}^{\mc L}{}_{\mc A}d\mc Z^{\mc A}{}_{\mc N}-d\bar{\mc Z}^{\mc L}{}_{\mc A}\mc Z^{\mc A}{}_{\mc N}\right)\G^{\mc N}{}_{\mc L}\\[0.2cm]
=&\frac{i}{2}(\bar{\mc Z}^c_{\mc A}d\mc Z^{\mc A}_c-d\bar{\mc Z}^c_{\mc A}\mc Z^{\mc A}_c)+\frac{i}{2}(\bar{\mc Z}_{\mc A\dot c}d\mc Z^{\mc A\dot c}-d\bar{\mc Z}_{\mc A\dot c}\mc Z^{\mc A\dot c})\\[0.2cm]
+&\frac{i}{2}(\bar\Psi^q_{\mc A}d\Psi^{\mc A}_q-d\bar\Psi^q_{\mc A}\Psi^{\mc A}_q)-\frac{i}{2}(\bar\Psi_{\mc A\dot q}d\Psi^{\mc A\dot q}-d\bar\Psi_{\mc A\dot q}\Psi^{\mc A\dot q})
\end{array}
\eeq
with the diagonal supermatrix
\beq
\G^{\mc N}{}_{\mc L}=\left(
\begin{array}{cc}
\g^{(0)\bs\nu}{}_{\bs\la} & 0 \\
0 & -\g^{(5)N}{}_L
\end{array}
\right)=\left(
\begin{array}{cc}
\begin{array}{cc}
-\de_a^b & 0 \\ 0 & \de^{\dot a}_{\dot b}
\end{array} & \bf0 \\[0.2cm]
\bf0 & \begin{array}{cc}
\de_q^p & 0 \\ 0 & -\de^{\dot q}_{\dot p}
\end{array}
\end{array}
\right).
\eeq
In its upper-diagonal block there is $\g^{(0)\bs\nu}{}_{\bs\la}$ matrix that enters kinetic term of the twistor Lagrangian for the bosonic particle (\ref{bos-twistor-action}). As discussed in section 2 it not merely breaks $SU(2,2)_R$ symmetry but actually 'switches' it to $SU(4)_R$ symmetry. In the case of superparticle supermatrix $\G$ interchanges $SU(2,2)_R$- and $SU(4)_R$-invariant blocks. So the supertwistor 1-form (\ref{kin-term-ph-vs-stwistor}) is invariant under 'twisted' $SU(2,2|4)_R$ symmetry with $SU(4)_R$ parameters in the upper-diagonal block and $SU(2,2)_R$ parameters in the lower-diagonal block. It can be made explicit by adopting another definition of the dual twistor supermatrix
\beq
\widetilde{\mc Z}^{\mc N}{}_{\mc A}=\G^{\mc N}{}_{\mc L}\bar{\mc Z}^{\mc L}{}_{\mc A}=\widetilde{\mc H}^{\mc N}{}_{\mc L}(\mc Z^{\mc B}{}_{\mc L})^\dagger\mc H^{\mc B}{}_{\mc A},
\eeq
where
\beq
\widetilde{\mc H}^{\mc N}{}_{\mc L}=\left(
\begin{array}{cc}
\de^{\bs\nu}_{\bs\la} & 0 \\
0 & -\g^{(5)N}{}_L
\end{array}
\right)
\eeq
resulting in the following expression for the 1-form (\ref{kin-term-ph-vs-stwistor})
\beq
\frac{i}{2}\left(\widetilde{\mc Z}^{\mc N}{}_{\mc A}d\mc Z^{\mc A}{}_{\mc N}-d\widetilde{\mc Z}^{\mc N}{}_{\mc A}\mc Z^{\mc A}{}_{\mc N}\right).
\eeq
This 'twisted' $SU(2,2|4)_R$ symmetry, however, will be broken by the supertwistor constraints, that we discuss below, similarly to breaking of the $SU(4)_R$ symmetry in the case of massive particle model in section 2. So let us write kinetic term of the superparticle's Lagrangian in the eight-supertwistor formulation in the form with manifest $(SU(2)_R)^4$ invariance
\beq\label{8-stwistor-kin}
\begin{array}{rl}
\mathscr L_{\mr{kin}}=&\frac{i}{2}(\bar{\mc Z}^c_{\mc A}\dot{\mc Z}^{\mc A}_c-\dot{\bar{\mc Z}}^c_{\mc A}\mc Z^{\mc A}_c)+\frac{i}{2}(\bar{\mc Z}_{\mc A\dot c}\dot{\mc Z}^{\mc A\dot c}-\dot{\bar{\mc Z}}_{\mc A\dot c}\mc Z^{\mc A\dot c})\\[0.2cm]
+&\frac{i}{2}(\bar\Psi^q_{\mc A}\dot\Psi^{\mc A}_q-\dot{\bar\Psi}^q_{\mc A}\Psi^{\mc A}_q)-\frac{i}{2}(\bar\Psi_{\mc A\dot q}\dot\Psi^{\mc A\dot q}-\dot{\bar\Psi}_{\mc A\dot q}\Psi^{\mc A\dot q}).
\end{array}
\eeq

To write down complete superparticle's Lagrangian in the supertwistor form we need to identify the set of constraints for supertwistors. To this end consider $SU(2,2|4)_L$-invariant  supermatrix quadratic in supertwistors
\beq\label{twistor-constr-smatrix}
\bar{\mc Z}^{\mc L}{}_{\mc A}\mc Z^{\mc A}{}_{\mc N}=
\left(
\begin{array}{cc}
\bar{\mc Z}^{\bs\la}_{\mc A}\mc Z^{\mc A}_{\bs\nu} & \bar{\mc Z}^{\bs\la}_{\mc A}\Psi^{\mc A}_{N} \\[0.2cm]
\bar\Psi^L_{\mc A}\mc Z^{\mc A}_{\bs\nu} & \bar\Psi^L_{\mc A}\Psi^{\mc A}_{N}
\end{array}
\right).
\eeq
Using the incidence relations (\ref{def-stwistor}) and (\ref{def-dual-stwistor}) it is easy to see that
\beq
\bar{\mc Z}^{\bs\la}_{\mc A}\Psi^{\mc A}_{N}=\bar\Psi^L_{\mc A}\mc Z^{\mc A}_{\bs\nu}\approx0
\eeq
constitute 32 odd constraints. From the relations for spinors (\ref{spin14-constr}) and (\ref{spin5-constr}) one finds that diagonal blocks of the supermatrix (\ref{twistor-constr-smatrix}) equal
\beq
\bar{\mc Z}^{\bs\la}_{\mc A}\mc Z^{\mc A}_{\bs\nu}=-\frac12
\left(
\begin{array}{cc}
\de^c_b(v^{\bs\al}_av^a_{\bs\al}) & 0 \\[0.2cm]
0 & \de^{\dot b}_{\dot c}(v^{\bs\al}_{\dot a}v^{\dot a}_{\bs \al})
\end{array}
\right),\quad
\bar\Psi^L_{\mc A}\Psi^{\mc A}_{N}=\frac12
\left(
\begin{array}{cc}
\de^q_p(\ell^A_r\ell^r_A) & 0 \\[0.2cm]
0 & -\de^{\dot p}_{\dot q}(\ell^A_{\dot r}\ell^{\dot r}_A)
\end{array}
\right).
\eeq
This allows to identify 28 bosonic constraints
\beq\label{su2-1}
L^a{}_b=\bar{\mc Z}^a_{\mc A}\mc Z^{\mc
A}_b-\frac12\de^a_b\bar{\mc Z}^c_{\mc A}\mc Z^{\mc A}_c\approx0,\quad
M_{\dot a}{}^{\dot b}=\bar{\mc Z}_{\mc A\dot a}\mc Z^{\mc
A\dot b}-\frac12\de_{\dot a}^{\dot b}\bar{\mc Z}_{\mc
A\dot c}\mc Z^{\mc A\dot c}\approx0,
\eeq
\beq\label{su22-grading-unity}
\bar{\mc Z}^a_{\mc A}\mc
Z^{\mc A\dot b}\approx0,\quad\bar{\mc Z}_{\mc A\dot b}\mc Z^{\mc
A}_{a}\approx0,
\eeq
\beq\label{su2-2}
R^{\,q}{}_p=\bar\Psi^q_{\mc A}\Psi^{\mc
A}_{p}-\frac12\de^q_p\bar\Psi^r_{\mc A}\Psi^{\mc A}_{r}\approx0,\quad
S_{\dot q}{}^{\dot p}=\bar\Psi_{\mc A\dot q}\Psi^{\mc A\dot
p}-\frac12\de_{\dot q}^{\dot p}\bar\Psi_{\mc A\dot r}\Psi^{\mc
A\dot r}\approx0,
\eeq
\beq\label{su4-grading-unity}
\bar\Psi^q_{\mc A}\Psi^{\mc A\dot
p}\approx0,\quad\bar\Psi_{\mc A\dot p}\Psi^{\mc A}_{q}\approx0.
\eeq
To identify other bosonic constraints consider D.B. relations
of the odd constraints. The form of the kinetic term
(\ref{8-stwistor-kin}) allows to find basic D.B. relations of
the supertwistors
\beq 
\begin{array}{c}
\{\mc Z^{\mc A}_a,\bar{\mc Z}^b_{\mc
B}\}_{D.B.}=i\de_a^b\de^{\mc A}_{\mc B},\quad\{\mc Z^{\mc A\dot
a},\bar{\mc Z}_{\mc B\dot b}\}_{D.B.}=i\de^{\dot
a}_{\dot b}\de^{\mc A}_{\mc B},\\[0.2cm] 
\{\Psi^{\mc A}_q,\bar\Psi^{p}_{\mc B}\}_{D.B.}=i\de_q^p\de^{\mc
A}_{\mc B},\quad\{\Psi^{\mc A\dot q},\bar\Psi_{\mc B\dot p}\}_{D.B.}=-i\de^{\dot q}_{\dot p}\de^{\mc A}_{\mc B}.
\end{array}
\eeq
With
these in mind we obtain necessary D.B. relations of the odd constraints
\beq\label{option1}
\begin{array}{c}
\{\bar{\mc Z}^b_{\mc A}\Psi^{\mc A}_{q},\bar\Psi^p_{\mc B}\mc Z^{\mc B}_{a}\}_{D.B.}=i\de^p_qL^b{}_a+i\de_a^bR^p{}_q+\frac{i}{2}\de^p_q\de^b_a(\bar{\mc Z}^c\mc Z_c+\bar\Psi^r\Psi_r),\\[0.2cm]
\{\bar{\mc Z}_{\mc A\dot b}\Psi^{\mc A\dot q},\bar\Psi_{\mc B\dot p}\mc Z^{\mc B\dot a}\}_{D.B.}=-i\de_{\dot p}^{\dot q}M_{\dot b}{}^{\dot a}+i\de_{\dot b}^{\dot a}S_{\dot p}{}^{\dot q}+\frac{i}{2}\de_{\dot p}^{\dot q}\de_{\dot b}^{\dot a}(-\bar{\mc Z}_{\dot c}\mc Z^{\dot c}+\bar\Psi_{\dot r}\Psi^{\dot r})
\end{array}
\eeq
and
\beq\label{option2}
\begin{array}{c}
\{\bar{\mc Z}^b_{\mc A}\Psi^{\mc A\dot q},\bar\Psi_{\mc B\dot p}\mc Z^{\mc B}_a\}_{D.B.}=-i\de_{\dot p}^{\dot q}L^b{}_a+i\de^b_aS_{\dot p}{}^{\dot q}+\frac{i}{2}\de_{\dot p}^{\dot q}\de^b_a(-\bar{\mc Z}^c\mc Z_c+\bar\Psi_{\dot r}\Psi^{\dot r}),\\[0.2cm]
\{\bar{\mc Z}_{\mc A\dot b}\Psi^{\mc A}_{q},\bar\Psi^{p}_{\mc B}\mc Z^{\mc B\dot a}\}_{D.B.}=i\de^p_qM_{\dot b}{}^{\dot a}+i\de_{\dot b}^{\dot a}R^p{}_q+\frac{i}{2}\de^{p}_q\de_{\dot b}^{\dot a}(\bar{\mc Z}_{\dot c}\mc Z^{\dot c}+\bar\Psi^r\Psi_r).
\end{array}
\eeq
There are two equivalent options to choose 16 odd first-class constraints of the model: those on the l.h.s. of (\ref{option1}) or (\ref{option2}). For definiteness let us consider odd constraints in (\ref{option1}) as the first-class ones. The consequence of such choice is that
\beq\label{29-30-constr}
\bar{\mc Z}^c\mc Z_c+\bar\Psi^r\Psi_r\approx0,\quad-\bar{\mc Z}_{\dot c}\mc Z^{\dot c}+\bar\Psi_{\dot r}\Psi^{\dot r}\approx0,
\eeq
while $-\bar{\mc Z}^c\mc Z_c+\bar\Psi_{\dot r}\Psi^{\dot r}$ and $\bar{\mc Z}_{\dot c}\mc Z^{\dot c}+\bar\Psi^r\Psi_r$ are non-zero since the constraints on the l.h.s. of (\ref{option2}) constitute 16 odd second-class constraints. Summing up constraints (\ref{29-30-constr}) and using the incidence relations (\ref{def-stwistor}), (\ref{def-dual-stwistor}) gives
\beq
\bar{\mc Z}^c\mc Z_c-\bar{\mc Z}_{\dot c}\mc Z^{\dot c}+\bar\Psi^r\Psi_r+\bar\Psi_{\dot r}\Psi^{\dot r}=v^{\bs\al}_cv^c_{\bs\al}+v^{\bs\al}_{\dot c}v^{\dot c}_{\bs\al}+\ell^A_r\ell^r_A-\ell^A_{\dot r}\ell^{\dot r}_A\approx0.
\eeq
The constraint on the r.h.s. is the square root of the mass-shell constraint (\ref{10d-mass-shell-constr}). This fixes the sign ambiguity in the definition of the square root of the mass-shell constraint. The difference of the constraints (\ref{29-30-constr}) is also the constraint
\beq\label{29-plus-30}
(\bar{\mc Z}^c\mc Z_c+\bar\Psi^r\Psi_r)-(-\bar{\mc Z}_{\dot c}\mc Z^{\dot c}+\bar\Psi_{\dot r}\Psi^{\dot r})=(\bar{\mc Z}_{\dot c}\mc Z^{\dot c}+\bar\Psi^r\Psi_r)-(-\bar{\mc Z}^c\mc Z_c+\bar\Psi_{\dot r}\Psi^{\dot r})\approx0,
\eeq
and equals the difference of $\bar{\mc Z}_{\dot c}\mc Z^{\dot c}+\bar\Psi^r\Psi_r$ and $-\bar{\mc Z}^c\mc Z_c+\bar\Psi_{\dot r}\Psi^{\dot r}$ so one can set
\beq
\bar{\mc Z}_{\dot c}\mc Z^{\dot c}+\bar\Psi^r\Psi_r=-\bar{\mc Z}^c\mc Z_c+\bar\Psi_{\dot r}\Psi^{\dot r}=Q\in\mathbb R\setminus\{0\}.
\eeq
As a result the products of supertwistors invariant under four different $SU(2)_R$ subgroups of $SU(2,2|4)_R$ 
can be parametrized as
\beq\label{norms}
\bar{\mc Z}^c\mc Z_c=-\frac{Q}{2}+q,\quad\bar\Psi^r\Psi_r=\frac{Q}{2}-q,\quad\bar{\mc Z}_{\dot c}\mc Z^{\dot c}=\bar\Psi_{\dot r}\Psi^{\dot r}=\frac{Q}{2}+q
\eeq
with an arbitrary real $q$. It is then easy to find another $Q$-independent linear combination of the supertwistor products
\beq\label{31st-constr}
\bar{\mc Z}^c\mc Z_c+\bar{\mc Z}_{\dot c}\mc Z^{\dot c}-\bar\Psi^r\Psi_r+\bar\Psi_{\dot r}\Psi^{\dot r}=4q.
\eeq
Requiring $q=0$ turns it into the 31st bosonic constraint. Constraints (\ref{29-plus-30}) and (\ref{31st-constr}) provide extra two conditions that being expressed in terms of the $Spin(1,4)$ and $Spin(5)$ variables balance the number of their independent components and those of $10d$ null momentum (see discussion after Eq.~(\ref{10d-mass-shell-constr})).

One can take another look at the above constraints viewing elements of the supermatrix (\ref{twistor-constr-smatrix}) as the generators of the 'twisted' $u(2,2|4)_R$ superalgebra. Then four $SU(2,2|4)_L$-invariant products of the $SU(2)_R$ doublets of supertwistors on the r.h.s. of the D.B. relations (\ref{option1}) and (\ref{option2}) can be parametrized as
\beq
\begin{array}{c}
\bar{\mc Z}^c\mc Z_c+\bar\Psi^r\Psi_r=\frac12(C+E+T)\approx0,\\[0.2cm]
\bar{\mc Z}_{\dot c}\mc Z^{\dot c}-\bar\Psi_{\dot r}\Psi^{\dot r}=-\frac12(C+E-T)\approx0, \\[0.2cm]
\bar{\mc Z}^c\mc Z_c-\bar\Psi_{\dot r}\Psi^{\dot r}=\frac12(C-E+T),\\[0.2cm]
\bar{\mc Z}_{\dot c}\mc Z^{\dot c}+\bar\Psi^r\Psi_r=-\frac12(C-E-T),
\end{array}
\eeq
where
\beq\label{e-plus-c}
C+E=\bar{\mc Z}^c\mc Z_c-\bar{\mc Z}_{\dot c}\mc Z^{\dot c}+\bar\Psi^r\Psi_r+\bar\Psi_{\dot r}\Psi^{\dot r}\approx0
\eeq
is the sum of the
$AdS$ energy (conformal dimension) generator $E=\bar\Psi^r\Psi_r+\bar\Psi_{\dot r}\Psi^{\dot r}$ and its counterpart in the $su(4)_R$ algebra $C=\bar{\mc Z}^c\mc Z_c-\bar{\mc Z}_{\dot c}\mc Z^{\dot c}$. Generator $T\in su(2,2|4)_R$ is
\beq\label{t}
T=\bar{\mc Z}^c\mc Z_c+\bar{\mc Z}_{\dot c}\mc Z^{\dot c}+\bar\Psi^r\Psi_r-\bar\Psi_{\dot r}\Psi^{\dot r}\approx0
\eeq
and the constraint (\ref{31st-constr}) equals the generator $U\in u(2,2|4)_R$.

After taking into account all of the constraints, the supermatrix (\ref{twistor-constr-smatrix}) is found to be proportional to the unit supermatrix
\beq
\bar{\mc Z}^{\mc L}{}_{\mc A}\mc Z^{\mc A}{}_{\mc N}\approx\frac{Q}{4}\de^{\mc L}_{\mc N}.
\eeq
Thus $\mc Z^{\mc A}{}_{\mc N}$ is an element of $U(2,2|4)\times\mathbb R$ modulo the gauge symmetries generated by the first-class constraints that we have to identify.

The choice of the first- and second-class fermionic constraints was made after Eq.~(\ref{option2}). There remains to consider D.B. relations of the bosonic constraints (\ref{su2-1})-(\ref{su4-grading-unity}), (\ref{31st-constr}), (\ref{e-plus-c}) and (\ref{t}) in order to find which of them are the first-class. $T\approx0$ D.B. commutes with all the constraints, whereas $U\approx0$ has non-zero D.B. relations only with odd constraints given in Appendix B and so they are the first-class constraints.
Four copies of the $su(2)$ generators (\ref{su2-1}) and (\ref{su2-2}) are also the first-class constraints, their D.B. relations with other constraints are given in Appendix B. Constraints (\ref{su22-grading-unity}) and (\ref{su4-grading-unity}) are the second-class ones as is seen from the D.B. relations
\beq
\begin{array}{c}
\{\bar{\mc Z}_{\mc A\dot a}\mc Z^{\mc A}_b,\bar{\mc Z}_{\mc B}^a\mc Z^{\mc B\dot b}\}_{D.B.}=i\de^a_bM_{\dot a}{}^{\dot b}-i\de_{\dot a}^{\dot b}L^a{}_b+\frac{i}{2}\de^a_b\de_{\dot a}^{\dot b}Q,\\[0.2cm]
\{\bar\Psi_{\mc A\dot q}\Psi^{\mc A}_p,\bar\Psi^q_{\mc B}\Psi^{\mc B\dot p}\}_{D.B.}=i\de^q_pS_{\dot q}{}^{\dot p}+i\de^{\dot p}_{\dot q}R^{\, q}{}_p+\frac{i}{2}\de^q_p\de_{\dot q}^{\dot p}Q.
\end{array}
\eeq
$E+C\approx0$ D.B. commutes with the first-class constraints. Its D.B. relations with the second-class constraints can be schematically presented as
\beq
\{E+C,G_{\pm}\}=\pm2iG_{\pm}, 
\eeq
where
\beq
G_{+}=\{\bar{\mc Z}_{\mc A}^a\mc Z^{\mc A\dot b},\bar\Psi_{\mc A}^q\Psi^{\mc A\dot p};\bar{\mc Z}^a_{\mc A}\Psi^{\mc A\dot q},\bar\Psi^q_{\mc A}\mc Z^{\mc A\dot a}\},\quad G_{-}=\{\bar{\mc Z}_{\mc A\dot a}\mc Z^{\mc A}_b,\bar\Psi_{\mc A\dot q}\Psi^{\mc A}_p;\bar\Psi_{\mc A\dot q}\mc Z_a^{\mc A},\bar{\mc Z}_{\mc A\dot a}\Psi^{\mc A}_q\}.
\eeq
Therefore we conclude that there are 15 bosonic first-class constraints that together with 16 fermionic first-class constraints generate $su(2|2)\oplus su(2|2)\oplus u(1)$ superalgebra of the gauge symmetry of the superparticle model in the eight-supertwistor formulation and
\beq
\mc Z^{\mc A}{}_{\mc N}\in U(2,2|4)\times\mr R/(SU(2|2)\times SU(2|2)\times U(1))
\eeq
having 18 bosonic and 16 fermionic independent components equal the number of physical degrees of freedom in the superspace formulation.

To conclude this section we present complete Lagrangian in the eight-supertwistor formulation
\beq
\mathscr L_{8-\mr{stwistor}}=\mathscr L_{\mr{kin}}+\mathscr L_{\mr{constr}},
\eeq
where the kinetic term is given in (\ref{8-stwistor-kin}) and the second summand is the linear combination of the constraints with Lagrange multipliers
\beq\label{lagr-constr}
\begin{array}{rl}
\mathscr L_{\mr{constr}}=&\La^b{}_aL^a{}_b+\La_{\dot b}{}^{\dot a}M_{\dot a}{}^{\dot b}+\La^p{}_qR^{\, q}{}_p+\La_{\dot p}{}^{\dot q}S_{\dot q}{}^{\dot p}+\La_{E+C}(E+C)+\La_TT+\La_UU\\[0.2cm]
+&i\La^a_p\bar\Psi^p\mc Z_a+i\bar\La^p_a\bar{\mc Z}^a\Psi_p+i\La_{\dot a}^{\dot p}\bar\Psi_{\dot p}\mc Z^{\dot a}+i\bar\La_{\dot p}^{\dot a}\bar{\mc Z}_{\dot a}\Psi^{\dot p}\\[0.2cm]
+&\La_{\dot ba}\bar{\mc Z}^a\mc Z^{\dot b}+\bar\La^{a\dot b}\bar{\mc Z}_{\dot b}\mc Z_a+\La_{\dot pq}\bar\Psi^q\Psi^{\dot p}+\bar\La^{q\dot p}\bar\Psi_{\dot p}\Psi_q\\[0.2cm]
+&i\La_{\dot ap}\bar\Psi^p\mc Z^{\dot a}
+i\bar\La^{p\,\dot a}\bar{\mc Z}_{\dot a}\Psi_p+i\La^{a\dot p}\bar\Psi_{\dot p}\mc Z_a+i\bar\La_{\dot p\, a}\bar{\mc Z}^a\Psi^{\dot p}.
\end{array}
\eeq 
Analysis of the conservation conditions of the constraints yields that the Lagrange multipliers for the second-class constraints in the two last lines in (\ref{lagr-constr}) turn to zero, while those for the first-class constraints in the first and second lines remain unfixed. 

\setcounter{equation}{0}
\section{Four-supertwistor formulation of massless superparticle on $\ads$ superbackground}

Reduction of the Lorentz-harmonic variables was discussed in section 3 and the results obtained there also apply to the $D=10$ superparticle model.
Superparticle's 10-momentum components tangent to $AdS_5$ can be brought to the form
\beq\label{reduced-5-momentum'}
p_{m'}=-\frac12v_{\bs\al}^b\g_{m'}{}^{\bs\al}{}_{\bs\bt}v_b^{\bs\bt}
\eeq
with
\beq
v_b^{\bs\al}v^b_{\bs\al}=-Q.
\eeq
Reduction of the $Spin(5)$ variables proceeds in the similar way.
Resulting expression for the momentum components tangent to $S^5$ is
\beq\label{s5-momentum-reduced}
p_{I'}=-\frac12\ell^{\, q}_B\g_{I'}{}^B{}_A\ell^A_q,
\eeq
where
\beq
\ell^A_q\ell^q_A=Q.
\eeq
Expressions for the momentum components (\ref{reduced-5-momentum'}) and (\ref{s5-momentum-reduced}) were the starting point to derive the four-supertwistor formulation of the massless superparticle on $AdS_5\times S^5$ superbackground from the first-order form of the superspace formulation \cite{U'18}. Resulting four-supertwistor representation of the superparticle's Lagrangian
\beq
\begin{array}{rl}
\mathscr L_{4-\mathrm{stwistor}}=&\frac{i}{2}\left(\bar{\mc Z}^a_{\mc A}\dot{\mc Z}^{\mc A}_a-\dot{\bar{\mc Z}}\vp{\bar{\mc Z}}^a_{\mc A}\mc Z^{\mc A}_a\right)+\frac{i}{2}(\bar\Psi^{q}_{\mc A}\dot\Psi^{\mc A}_{q}-\dot{\bar\Psi}^{q}_{\mc A}\Psi^{\mc A}_{q}) \\[0.2cm]
+&\Lambda^b{}_aL^a{}_b+\Lambda^p{}_qR^{\, q}{}_p+\Lambda(\bar{\mc Z}_{\mc A}^{a}\mc Z^{\mc A}_{a}+\bar\Psi_{\mc A}^{q}\Psi^{\mc A}_{q}) \\[0.2cm]
+&i\Lambda^a_{q}\bar\Psi^{q}_{\mc A}\mc Z^{\mc A}_a+i\bar\Lambda^{q}_a\bar{\mc Z}^a_{\mc A}\Psi^{\mc A}_{q}
\end{array}
\eeq 
coincides with that obtained in Ref.~\cite{Bars-ads5s5} by
partial gauge fixing $2T$ superparticle model in $2+10$
dimensions. There are seven bosonic 
\beq\label{bose-constr-4tw}
L^a{}_b=\bar{\mc Z}^a_{\mc A}\mc Z^{\mc
A}_b-\frac12\de^a_b\bar{\mc Z}^c_{\mc A}\mc Z^{\mc
A}_c\approx0,\quad R^{\,q}{}_p=\bar\Psi^q_{\mc A}\Psi^{\mc
A}_{p}-\frac12\de^q_p\bar\Psi^r_{\mc A}\Psi^{\mc
A}_{r}\approx0,\quad \bar{\mc Z}_{\mc A}^a\mc Z^{\mc A}_a+\bar\Psi^q_{\mc A}\Psi^{\mc A}_q\approx0
\eeq 
and eight fermionic 
\beq\label{fermi-constr-4tw}
\bar\Psi^{q}_{\mc A}\mc Z^{\mc A}_a\approx0,\quad\bar{\mc
Z}^a_{\mc A}\Psi^{\mc A}_{q}\approx0 
\eeq 
first-class constraints
quadratic in supertwistors. These are the generators of the
$su(2|2)$ gauge superalgebra named color superalgebra in
\cite{Bars-twistor-string}.

In our previous work \cite{U'18} massless superparticle model in the four-supertwistor formulation was quantized in the simplest case when it propagates only within the $AdS_5$ subspace of the $AdS_5\times S^5$ superspace, i.e. momentum components in directions tangent to $S^5$ vanish. In supertwistor formulation this amounts to vanishing of $\Psi^{\mc A}_q$ and $\bar\Psi^q_{\mc A}$ supertwistors so that only the c-type supertwistors $\mc Z^{\mc A}_a$ and $\bar{\mc Z}^a_{\mc A}$ are dynamical variables. It was demonstrated that the states of quantized superparticle coincide with the $D=5$ $N=8$ gauged supergravity multiplet \cite{GM'85} both in the supertwistor and superoscillator approaches.

In general all four supertwistors contribute. Their
(anti)commutation relations are found to be \beq [\mc Z^{\mc
A}_a,\bar{\mc Z}^b_{\mc B}\}=\de_a^b\de^{\mc A}_{\mc
B},\quad[\Psi^{\mc A}_q,\bar\Psi^{p}_{\mc B}\}=\de_q^p\de^{\mc
A}_{\mc B}. 
\eeq 
To set up the stage for superambitwistor
quantization let us introduce individual notation for the
components of $c$- and $a$-type supertwistors \beq
\begin{array}{c}
\mc Z^{\mc A}_1
\equiv
\mc Z^{\mc A}=\left(
\begin{array}{c}
Z^{\boldsymbol{\al}} \\ \eta^A
\end{array}
\right),
\quad
\bar{\mc Z}^1_{\mc A}\equiv
\bar{\mc Z}_{\mc A}=(\bar Z_{\boldsymbol{\al}},\:\bar\eta_A),
\\[0.4cm]
\mc Z^{\mc A}_2
\equiv
\mc W^{\mc A}=
\left(
\begin{array}{c}
W^{\boldsymbol{\al}} \\ \zeta^A
\end{array}
\right),\quad \bar{\mc Z}^2_{\mc A}\equiv\bar{\mc W}_{\mc A}
=(\bar W_{\boldsymbol{\al}},\:\bar\zeta_A)
\end{array}
\eeq
and
\beq
\begin{array}{c}
\Psi^{\mc A}_1
\equiv
\Psi^{\mc A}=\left(
\begin{array}{c}
\xi^{\boldsymbol{\al}} \\ L^A
\end{array}
\right),
\quad
\bar\Psi^1_{\mc A}\equiv
\bar\Psi_{\mc A}=(\bar\xi_{\boldsymbol{\al}},\:\bar L_A),
\\[0.4cm]
\Psi^{\mc A}_2
\equiv
\Xi^{\mc A}=
\left(
\begin{array}{c}
\varrho^{\boldsymbol{\al}} \\ M^A
\end{array}
\right),\quad\bar\Psi^2_{\mc A}\equiv\bar\Xi_{\mc A}
=(\bar\varrho_{\boldsymbol{\al}},\:\bar M_A).
\end{array}
\eeq
Let us further realize components of quantized supertwistors as the multiplication and differentiation operators
\beq\label{qtwistor-realization}
\begin{array}{c}
Z^{\boldsymbol{\al}}\rightarrow Z^{\boldsymbol{\al}},\quad\bar Z_{\boldsymbol{\al}}\rightarrow-\frac{\pt}{\pt\vp{\bar Z^{\boldsymbol{\al}}} Z^{\boldsymbol{\al}}},\quad\eta^A\rightarrow\eta^A,\quad\bar\eta_A\rightarrow\frac{\vec\pt}{\pt\vp{\hat\eta^A}\eta^A}, \\[0.2cm]
W^{\boldsymbol{\al}}\rightarrow\frac{\pt}{\pt\vp{\hat W_{\boldsymbol{\al}}}\bar W_{\boldsymbol{\al}}},\quad\bar W_{\boldsymbol{\al}}\rightarrow\bar W_{\boldsymbol{\al}},\quad\zeta^A\rightarrow\frac{\vec\pt}{\pt\vp{\hat\zeta_A}\bar\zeta_A},\quad\bar\zeta_A\rightarrow\bar\zeta_A
\end{array}
\eeq
and
\beq\label{qtwistor-realization2}
\begin{array}{c}
\xi^{\boldsymbol{\al}}\rightarrow \xi^{\boldsymbol{\al}},\quad\bar\xi_{\boldsymbol{\al}}\rightarrow\frac{\vec\pt}{\pt\xi^{\boldsymbol{\al}}},\quad L^A\rightarrow L^A,\quad\bar L_A\rightarrow-\frac{\pt}{\pt\vp{\hat L^A}L^A}, \\[0.2cm]
\varrho^{\boldsymbol{\al}}\rightarrow\frac{\vec\pt}{\pt\vp{\hat\varrho_{\boldsymbol{\al}}}\bar\varrho_{\boldsymbol{\al}}},\quad\bar\varrho_{\boldsymbol{\al}}\rightarrow\bar\varrho_{\boldsymbol{\al}},\quad M^A\rightarrow\frac{\pt}{\pt\vp{\hat M_A}\bar M_A},\quad\bar M_A\rightarrow\bar M_A,
\end{array}
\eeq 
where odd derivatives are defined to act from the left. Then
the  constraints (\ref{bose-constr-4tw}) and
(\ref{fermi-constr-4tw}) become differential operators acting in
superambitwistor space. Let us a la Gupta-Bleuler quantization approach 
assume that the superparticle's wave function is annihilated by
the diagonal elements of the bosonic constraint matrices
$L^1{}_1=-L^2{}_2\approx0$ and $R^1{}_1=-R^2{}_2\approx0$, as well
as off-diagonal elements $L^2{}_1\approx0$ and
$R^2{}_1\approx0$  that are given by the differential operators of
utmost the first-order. As a result the superparticle's wave
function satisfies the following equations 
\beq\label{4-bos-eq}
\begin{array}{c}
(H_{\mc Z}+H_{\bar{\mc W}})\mathscr F_{(h_{\mc Z},\, h_{\bar{\mc W}}|h_{\Psi},\, h_{\bar\Xi})}(\mc Z,\bar{\mc W},\Psi,\bar\Xi)=0,\quad
(H_{\Psi}+H_{\bar\Xi})\mathscr F_{(h_{\mc Z},\, h_{\bar{\mc W}}|h_{\Psi},\, h_{\bar\Xi})}(\mc Z,\bar{\mc W},\Psi,\bar\Xi)=0,\\[0.2cm]
\bar{\mc W}_{\mc A}\mc Z^{\mc A}\mathscr F_{(h_{\mc Z},\, h_{\bar{\mc W}}|h_{\Psi},\, h_{\bar\Xi})}(\mc Z,\bar{\mc W},\Psi,\bar\Xi)=\bar\Xi_{\mc A}\Psi^{\mc A}\mathscr F_{(h_{\mc Z},\, h_{\bar{\mc W}}|h_{\Psi},\, h_{\bar\Xi})}(\mc Z,\bar{\mc W},\Psi,\bar\Xi)=0,
\end{array}
\eeq 
where 
\beq H_{\mc Z}=\mc Z^{\mc A}\frac{\pt}{\pt\mc Z^{\mc
A}}=Z^{\bs\al}\frac{\pt}{\pt
Z^{\bs\al}}+\eta^A\frac{\pt}{\pt\eta^A},\quad H_{\bar{\mc
W}}=\bar{\mc W}_{\mc A}\frac{\pt}{\pt\vp{\hat{\mc W}_{\mc
A}}\bar{\mc W}_{\mc A}}=\bar W_{\bs\al}\frac{\pt}{\pt\vp{\hat
W_{\bs\al}}\bar
W_{\bs\al}}+\bar\zeta_A\frac{\pt}{\pt\vp{\hat\zeta_A}\bar\zeta_A}
\eeq 
and 
\beq H_{\Psi}=\Psi^{\mc A}\frac{\pt}{\pt\Psi^{\mc
A}}=\xi^{\bs\al}\frac{\pt}{\pt\xi^{\bs\al}}+L^A\frac{\pt}{\pt
L^A},\quad H_{\bar\Xi}=\bar\Xi_{\mc A}\frac{\pt}{\pt\vp{\hat\Xi_{\mc
A}}\bar\Xi_{\mc
A}}=\bar\varrho_{\bs\al}\frac{\pt}{\pt\vp{\hat\varrho_{\bs\al}}\bar\varrho_{\bs\al}}+\bar
M_A\frac{\pt}{\pt\vp{\hat M_A}\bar M_A} 
\eeq 
are dilatation
operators in each of the ambitwistor variables. Their eigenvalues
-- homogeneity degrees in respective arguments $h_{\mc Z}$,
$h_{\bar{\mc W}}$, $h_{\Psi}$ and $h_{\bar\Xi}$ are indicated in
the subscript of the superparticle's wave function. Extra equation
\beq 
(H_{\mc Z}-H_{\bar{\mc W}}+H_{\Psi}-H_{\bar\Xi})\mathscr
F_{(h_{\mc Z},\, h_{\bar{\mc W}}|h_{\Psi},\, h_{\bar\Xi})}(\mc
Z,\bar{\mc W},\Psi,\bar\Xi)=0 
\eeq 
comes from the last constraint
in (\ref{bose-constr-4tw}). Thus there are three equations for
four homogeneity degrees so we can introduce one arbitrary
parameter $h$ and write $\mathscr F_{(h_{\mc Z},\, h_{\bar{\mc
W}}|h_{\Psi},\, h_{\bar\Xi})}(\mc Z,\bar{\mc
W},\Psi,\bar\Xi)=\mathscr F_{(h,\,-h|-h,\, h)}(\mc Z,\bar{\mc
W},\Psi,\bar\Xi)$. Besides that impose on the superparticle's wave function 
the equations that stem from the odd constraints and
are given by the utmost first-order differential operators. In the realization (\ref{qtwistor-realization}), (\ref{qtwistor-realization2}) two odd constraints $\bar{\mc W}_{\mc A}\Psi^{\mc A}\approx0$ and $\bar\Xi_{\mc A}\mc Z^{\mc A}\approx0$ translate into algebraic 
equations 
\beq 
\bar{\mc W}_{\mc A}\Psi^{\mc A}\mathscr F_{(h,\,-h|-h,\, h)}(\mc Z,\bar{\mc W},\Psi,\bar\Xi)=\bar\Xi_{\mc A}\mc Z^{\mc A}\mathscr F_{(h,\,-h|-h,\, h)}(\mc Z,\bar{\mc W},\Psi,\bar\Xi)=0 
\eeq 
that together with two algebraic equations in (\ref{4-bos-eq}) can be taken into account by introducing the delta-function factors
\beq\label{delta-ansatz}
\mathscr F_{(h,\,-h|-h,\, h)}(\mc Z,\bar{\mc W},\Psi,\bar\Xi)=\de(\bar{\mc W}\mc Z)\de(\bar\Xi\Psi)\de(\bar{\mc W}\Psi)\de(\bar\Xi\mc Z)F_{(h,\,-h|-h,\, h)}(\mc Z,\bar{\mc W},\Psi,\bar\Xi).
\eeq
Additionally two pairs of the conjugate constraints $\bar\Psi_{\mc A}\mc Z^{\mc A}\approx0$, $\bar{\mc Z}_{\mc A}\Psi^{\mc A}\approx0$ and $\bar\Xi_{\mc A}\mc W^{\mc A}\approx0$, $\bar{\mc W}_{\mc A}\Xi^{\mc A}\approx0$ yield first-order differential equations. Assuming that in each pair of the equations associated with these constraints one equation is imposed on the superparticle's wave
function and another on its conjugate gives four possible sets of equations that will be analyzed elsewhere. As an example we present one set of such equations 
\beq\label{4-odd-eq}
(-)^{\mc A}\mc Z^{\mc A}\frac{\pt}{\pt\Psi^{\mc A}}\mathscr F_{(h,\,-h|-h,\, h)}(\mc Z,\bar{\mc W},\Psi,\bar\Xi)=\bar\Xi_{\mc A}\frac{\pt}{\pt\vp{\hat{\mc W}^{\mc A}}\bar{\mc W}^{\mc A}}\mathscr F_{(h,\,-h|-h,\, h)}(\mc Z,\bar{\mc W},\Psi,\bar\Xi)=0,  
\eeq 
where in the first equation it is assumed that $SU(2,2)_L$ index has parity zero and $SU(4)_L$ index has parity one. Since the differential operators in (\ref{4-odd-eq}) annihilate delta-function factors introduced in (\ref{delta-ansatz}) we obtain 
\beq
(-)^{\mc A}\mc Z^{\mc A}\frac{\pt}{\pt\Psi^{\mc A}}F_{(h,\,-h|-h,\, h)}(\mc Z,\bar{\mc W},\Psi,\bar\Xi)=\bar\Xi_{\mc A}\frac{\pt}{\pt\vp{\hat{\mc W}^{\mc A}}\bar{\mc W}^{\mc A}}F_{(h,\,-h|-h,\, h)}(\mc Z,\bar{\mc W},\Psi,\bar\Xi)=0. 
\eeq
Analysis of these and other equations and of the Penrose transform of the function $F_{(h,\,-h|-h,\, h)}(\mc Z,\bar{\mc W},\Psi,\bar\Xi)$ is postponed for future study.

\section{Conclusion}

In the present paper we addressed the issue of deriving
supertwistor formulation for $D=10$ massless superparticle on $\ads$
superbackground starting with the first-order representation of
the superspace Lagrangian, in which momentum components tangent to
$AdS_5$ and $S^5$ are expressed via the $Spin(1,4)$ and $Spin(5)$
variables. Resulting supertwistors coincide with those
found in \cite{Bars-twistor-string} so that we not only
established their origin but also derived the incidence relations
with the $\ads$ superspace coordinates via the $\coset$ supercoset representative. Using
the incidence relations we have identified the set of bosonic and
fermionic constraints that these supertwistors satisfy and
calculated their classical algebra. It appears that among the
constraints there are the second-class ones that complicates
canonical analysis of the model. We have shown that using the
constraints for the spinor variables it is possible to exclude half of them 
from the expressions for the momentum
components reducing the eight-supertwistor form of the
superparticle's Lagrangian to the four-supertwistor form.
Respective supertwistors were proposed in \cite{Bars-ads5s5} and the
incidence relations with the $\ads$ coordinates were obtained in
\cite{U'18}. This not only explains the origin of both kinds of
supertwistors pertinent to the $\ads$ superparticle model but also
establishes the relation between them. The advantage of the
four-supertwistor formulation is the presence of only the
first-class constraints. As a result we have obtained the set of
equations for the superparticle's wave function in the
superambitwistor space that generalize those of Ref.~\cite{U'18}.

In general the distinctive feature of the twistor formulations is that the Lagrangian of the superparticle models is quadratic in supertwistors as well as the constraints. This is of particular importance for curved superbackgrounds such as $\ads$ one, for which, depending on the parametrization, supervielbein components may be highly non-linear in the superspace coordinates. The difficulty with the supertwistor formulation is to provide a space-time interpretation of the states of quantized model. In the case when the superparticle moves within the $AdS_5$ subspace of the $\ads$ superspace we succeeded in \cite{U'18} to map its quantum states to those of the supermultiplet of $D=5$ $N=8$ gauged supergravity \cite{GM'85} both in the superoscillator and supertwistor approaches. Next task is to extend these results to the generic case and find a relation between the components of the superparticle's wave function and the towers of Kaluza-Klein states of the IIB supergravity compactified on $\ads$ superbackground.

We also plan to examine the spectrum of quantized tensionless supertwistor string on $\ads$ sketched in \cite{U'18} and its relation to the higher-spin supermultiplets emerging on both sides of $AdS_5/CFT_4$ duality in the limit of vanishing 't Hooft coupling \cite{Bo}, \cite{Witten-talk}. Supertwistor formulation of the tensile $\ads$ superstring is another direction of the generalization of the results reported here. Such a formulation could be an interesting alternative to the superspace formulation based on the $\coset$ supercoset sigma-model \cite{MT'98}, \cite{Rahmfeld}.

Finally in \cite{Bars-ads5s5} there were considered twistor gauges
for superparticle models relevant to other maximally
supersymmetric backgrounds. It is of interest to obtain incidence
relations for these supertwistors starting with the superspace
formulations of conventional superparticle models on such
superbackgrounds and study their quantization.

\appendix
\setcounter{equation}{0}
\section{Details of the spinor algebra}

In Appendix A of Ref.~\cite{U'18} there were collected the details of the spinor algebra in $D=2+4$ and $D=1+4$ dimensions alongside with the realization of the $D=1+4$ $\g$-matrices with manifest $SL(2,\mathbb C)$ symmetry that was used for the supermatrix realization of the generators of $psu(2,2|4)$ superalgebra as $D=4$ $N=4$ superconformal algebra. In order to spell out the relations between the $D=1+4$ spinor and vector Lorentz harmonics parametrizing the $SO(1,4)/SO(4)$ coset space another realization of the $D=1+4$ $\g$-matrices is used that manifests $SO(4)\simeq SU(2)\times SU(2)$ symmetry
\beq\label{g-realiz}
\g^{0}{}_{\bs\al}{}^{\bs\bt}=\left(
\begin{array}{cc}
\de_a^b & 0 \\
0 & -\de^{\dot a}_{\dot b}
\end{array}
\right), \quad
\g^{\hat I}{}_{\bs\al}{}^{\bs\bt}=\left(
\begin{array}{cc}
0 & \s^{\hat I}_{a\dot b} \\
\tilde\s^{\hat I\dot ab} & 0
\end{array}
\right),\quad\hat I=1,2,3,4. 
\eeq 
$D=1+4$ $\g$-matrix algebra 
\beq
\g^{m'}{}_{\bs\al}{}^{\bs\bt}\g^{n'}{}_{\bs\bt}{}^{\bs\de}+\g^{n'}{}_{\bs\al}{}^{\bs\bt}\g^{m'}{}_{\bs\bt}{}^{\bs\de}=-2\eta^{m'n'}\de_{\bs\al}^{\bs\de},\quad
m'=0,\hat I 
\eeq 
in this realization is fulfilled due to relations
that obey $\s^{\hat I}_{a\dot b}$ and $\tilde\s^{\hat I\dot
ab}=\varepsilon^{bc}\varepsilon^{\dot a\dot d}\s^{\hat I}_{c\dot
d}$ 
\beq\label{d4-sigma-matrices} 
\s^{\hat I}_{a\dot
b}\tilde\s^{\hat J\dot bc}+\s^{\hat J}_{a\dot b}\tilde\s^{\hat
I\dot bc}=-2\de^{\hat I\hat J}\de_a^c. 
\eeq 
Matrices $\s^{1,2,3}$ coincide
with the Pauli matrices and $\s^4=\pm i\,\mr I$. Charge conjugation
matrices is this realization are chosen as 
\beq
C_{\bs\al\bs\bt}=\left(
\begin{array}{cc}
-\varepsilon_{ab} & 0 \\
0 & \varepsilon^{\dot a\dot b}
\end{array}
\right), \quad
C^{\bs\al\bs\bt}=\left(
\begin{array}{cc}
-\varepsilon^{ab} & 0 \\
0 & \varepsilon_{\dot a\dot b}
\end{array}
\right),
\eeq
where $\varepsilon_{ab}$, $\varepsilon^{ab}$: $\varepsilon_{ab}\varepsilon^{bc}=\de^c_a$ and $\varepsilon_{\dot a\dot b}$, $\varepsilon^{\dot a\dot b}$: $\varepsilon_{\dot a\dot b}\varepsilon^{\dot b\dot c}=\de_{\dot a}^{\dot c}$ are used to change the positions of the fundamental representation indices of the two $SU(2)$ factors of $SO(4)$. $\g$-matrix realization (\ref{g-realiz}) and that used in Ref.~\cite{U'18} are related by the similarity transformation analogous to that connecting $D=1+3$ $\g$-matrices in canonical and Weyl bases (see, e.g. \cite{WB}).

To write expressions for the $D=5$ vector and spinor harmonics we use the following realization of $D=5$ $\g$-matrices with manifest $SO(4)$ covariance
\beq
\g^{\hat I'}{}_{L}{}^{N}=\left(
\begin{array}{cc}
0 & \s^{\hat I'}_{q\dot p} \\
-\tilde\s^{\hat I'\dot qp} & 0
\end{array}
\right),\quad\hat I'=1,2,3,4,\quad
\g^5{}_{L}{}^{N}=\left(
\begin{array}{cc}
\de_q^p & 0 \\
0 & -\de^{\dot q}_{\dot p}
\end{array}
\right),
\eeq
where $\s^{\hat I'}_{q\dot p}$ and $\tilde\s^{\hat I'\dot qp}$ satisfy the same relations as in (\ref{d4-sigma-matrices}). $Spin(5)\simeq USp(4)$ charge conjugation matrices in this realization equal 
\beq 
C_{LN}=\left(
\begin{array}{cc}
\varepsilon_{qp} & 0 \\
0 & \varepsilon^{\dot q\dot p}
\end{array}
\right),\quad 
C^{LN}=\left(
\begin{array}{cc}
\varepsilon^{qp} & 0 \\
0 & \varepsilon_{\dot q\dot p}
\end{array}
\right). 
\eeq 

\setcounter{equation}{0}
\section{Algebra of the constraints for eight-supertwistor formulation}

For completeness we present here those D.B. relations of the constraints in the eight-supertwistor formulation of the superparticle that have not been given in the main text.

D.B. relations of the fermionic constraints apart from those in Eqs.~(\ref{option1}) and (\ref{option2}) are
\beq
\begin{array}{c}
\{\bar{\mc Z}^a_{\mc A}\Psi^{\mc A}_q,\bar\Psi_{\mc B\dot p}\mc Z^{\mc B}_b\}_{D.B.}=i\de^a_b\bar\Psi_{\dot p}\Psi_q\approx0,\quad
\{\bar{\mc Z}^a_{\mc A}\Psi^{\mc A}_q,\bar\Psi^p_{\mc B}\mc Z^{\mc B\dot b}\}_{D.B.}=i\de^p_q\bar{\mc Z}^a\mc Z^{\dot b}\approx0,\\[0.2cm]
\{\bar{\mc Z}^a_{\mc A}\Psi^{\mc A\dot q},\bar\Psi^p_{\mc B}\mc Z^{\mc B}_b\}_{D.B.}=i\de^a_b\bar\Psi^p\Psi^{\dot q}\approx0,\quad
\{\bar{\mc Z}^a_{\mc A}\Psi^{\mc A\dot q},\bar\Psi_{\mc B\dot p}\mc Z^{\mc B\dot b}\}_{D.B.}=-i\de_{\dot p}^{\dot q}\bar{\mc Z}^a\mc Z^{\dot b}\approx0,\\[0.2cm]
\{\bar{\mc Z}_{\mc A\dot a}\Psi^{\mc A}_q,\bar\Psi^p_{\mc B}\mc Z^{\mc B}_b\}_{D.B.}=i\de^p_q\bar{\mc Z}_{\dot a}\mc Z_b\approx0,\quad
\{\bar{\mc Z}_{\mc A\dot a}\Psi^{\mc A}_q,\bar\Psi_{\mc B\dot p}\mc Z^{\mc B\dot b}\}_{D.B.}=i\de_{\dot a}^{\dot b}\bar\Psi_{\dot p}\Psi_q\approx0,\\[0.2cm]
\{\bar{\mc Z}_{\mc A\dot a}\Psi^{\mc A\dot q},\bar\Psi_{\mc B\dot p}\mc Z^{\mc B}_b\}_{D.B.}=-i\de_{\dot p}^{\dot q}\bar{\mc Z}_{\dot a}\mc Z_b\approx0,\quad
\{\bar{\mc Z}_{\mc A\dot a}\Psi^{\mc A\dot q},\bar\Psi^p_{\mc B}\mc Z^{\mc B\dot b}\}_{D.B.}=i\de_{\dot a}^{\dot b}\bar\Psi^p\Psi^{\dot q}\approx0.
\end{array}
\eeq

D.B. relations of $U$ with the fermionic constraints can be schematically written as
\beq
\{U,Q_{\pm}\}_{D.B.}=\pm2iQ_{\pm},
\eeq
where
\beq
Q_+=\{\bar{\mc Z}^a_{\mc A}\Psi^{\mc A}_q,\bar{\mc Z}_{\mc A\dot a}\Psi^{\mc A\dot q},\bar{\mc Z}_{\mc A\dot a}\Psi^{\mc A}_q,\bar{\mc Z}^a_{\mc A}\Psi^{\mc A\dot q}\},\quad Q_-=\{\bar\Psi^q_{\mc A}\mc Z^{\mc A}_a,\bar\Psi_{\mc A\dot q}\mc Z^{\mc A\dot a},\bar\Psi^q_{\mc A}\mc Z^{\mc A\dot a},\bar\Psi_{\mc A\dot q}\mc Z^{\mc A}_a\}.
\eeq

D.B. relations of the constraints (\ref{su2-1}) and (\ref{su22-grading-unity}) are
\beq
\begin{array}{c}
\{L^a{}_b,\bar{\mc Z}_{\mc A\dot a}\mc Z^{\mc A}_c\}_{D.B.}=\frac{i}{2}\de^a_b\bar{\mc Z}_{\mc A\dot a}\mc Z^{\mc A}_c-i\de^a_c\bar{\mc Z}_{\mc A\dot a}\mc Z^{\mc A}_b,\quad
\{L^a{}_b,\bar{\mc Z}_{\mc A}^c\mc Z^{\mc A\dot b}\}_{D.B.}=-\frac{i}{2}\de^a_b\bar{\mc Z}_{\mc A}^c\mc Z^{\mc A\dot b}+i\de^c_b\bar{\mc Z}_{\mc A}^a\mc Z^{\mc A\dot b},\\[0.2cm]
\{M_{\dot a}{}^{\dot b},\bar{\mc Z}_{\mc A\dot c}\mc Z^{\mc A}_b\}_{D.B.}=-\frac{i}{2}\de_{\dot a}^{\dot b}\bar{\mc Z}_{\mc A\dot c}\mc Z^{\mc A}_b+i\de_{\dot c}^{\dot b}\bar{\mc Z}_{\mc A\dot a}\mc Z^{\mc A}_b,\quad
\{M_{\dot a}{}^{\dot b},\bar{\mc Z}_{\mc A}^a\mc Z^{\mc A\dot c}\}_{D.B.}=\frac{i}{2}\de_{\dot a}^{\dot b}\bar{\mc Z}_{\mc A}^a\mc Z^{\mc A\dot c}-i\de^{\dot c}_{\dot a}\bar{\mc Z}_{\mc A}^a\mc Z^{\mc A\dot b},\\[0.2cm]
\{L^a{}_b,L^c{}_d\}_{D.B.}=i(\de^c_bL^a{}_d-\de^a_dL^c{}_b),\quad
\{M_{\dot a}{}^{\dot b},M_{\dot c}{}^{\dot d}\}_{D.B.}=i(\de^{\dot b}_{\dot c}M_{\dot a}{}^{\dot d}-\de_{\dot a}^{\dot d}M_{\dot c}{}^{\dot b}).
\end{array}
\eeq
Similarly D.B. relations of the constraints (\ref{su2-2}) and (\ref{su4-grading-unity}) read
\beq
\begin{array}{c}
\{R^{\, q}{}_p,\bar\Psi_{\mc A\dot q}\Psi^{\mc A}_r\}_{D.B.}=\frac{i}{2}\de^q_p\bar\Psi_{\mc A\dot q}\Psi^{\mc A}_r-i\de^q_r\bar\Psi_{\mc A\dot q}\Psi^{\mc A}_p,\quad
\{R^{\, q}{}_p,\bar\Psi^r_{\mc A}\Psi^{\mc A\dot q}\}_{D.B.}=-\frac{i}{2}\de^q_p\bar\Psi^r_{\mc A}\Psi^{\mc A\dot q}+i\de^r_p\bar\Psi^q_{\mc A}\Psi^{\mc A\dot q},\\[0.2cm]
\{S_{\dot q}{}^{\dot p},\bar\Psi_{\mc A\dot r}\Psi^{\mc A}_q\}_{D.B.}=\frac{i}{2}\de_{\dot q}^{\dot p}\bar\Psi_{\mc A\dot r}\Psi^{\mc A}_q-i\de^{\dot p}_{\dot r}\bar\Psi_{\mc A\dot q}\Psi^{\mc A}_q,\quad
\{S_{\dot q}{}^{\dot p},\bar\Psi^q_{\mc A}\Psi^{\mc A\dot r}\}_{D.B.}=-\frac{i}{2}\de_{\dot q}^{\dot p}\bar\Psi^q_{\mc A}\Psi^{\mc A\dot r}+i\de^{\dot r}_{\dot q}\bar\Psi^q_{\mc A}\Psi^{\mc A\dot p},\\[0.2cm]
\{R^q{}_p,R^s{}_t\}_{D.B.}=i(\de^s_pR^q{}_t-\de^q_tR^s{}_p),\quad\{S_{\dot q}{}^{\dot p},S_{\dot s}{}^{\dot t}\}_{D.B.}=i(\de^{\dot t}_{\dot q}S_{\dot s}{}^{\dot p}-\de^{\dot p}_{\dot s}S_{\dot q}{}^{\dot t}).
\end{array}
\eeq

Finally we present D.B. relations for bosonic and fermionic
constraints. D.B. relations that involve fermionic first-class
constraints are 
\beq
\begin{array}{c}
\{\bar{\mc Z}_{\mc A\dot c}\mc Z^{\mc A}_b,\bar{\mc Z}^a_{\mc B}\Psi^{\mc B}_q\}_{D.B.}=i\de^a_b\bar{\mc Z}_{\mc A\dot c}\Psi^{\mc A}_q,\quad
\{L^c{}_b,\bar{\mc Z}^a_{\mc A}\Psi^{\mc A}_q\}_{D.B.}=-\frac{i}{2}\de^c_b\bar{\mc Z}^a_{\mc A}\Psi^{\mc A}_q+i\de^a_b\bar{\mc Z}^c_{\mc A}\Psi^{\mc A}_q, \\[0.2cm]
\{\bar\Psi^r_{\mc A}\Psi^{\mc A\dot p},\bar{\mc Z}^a_{\mc B}\Psi^{\mc B}_q\}_{D.B.}=-i\de^r_p\bar{\mc Z}^a_{\mc A}\Psi^{\mc A\dot p},\quad
\{R^{\, r}{}_p,\bar{\mc Z}^a_{\mc A}\Psi^{\mc A}_q\}_{D.B.}=\frac{i}{2}\de^r_p\bar{\mc Z}^a_{\mc A}\Psi^{\mc A}_q-i\de^r_q\bar{\mc Z}^a_{\mc A}\Psi^{\mc A}_p; \\[0.2cm]
\{\bar{\mc Z}^c_{\mc A}\mc Z^{\mc A\dot b},\bar\Psi^q_{\mc B}\mc Z^{\mc B}_a\}_{D.B.}=-i\de^c_a\bar\Psi^q_{\mc A}\mc Z^{\mc A\dot b},\quad
\{L^c{}_b,\bar\Psi^q_{\mc A}\mc Z^{\mc A}_a\}_{D.B.}=\frac{i}{2}\de^c_b\bar\Psi^q_{\mc A}\mc Z^{\mc A}_a-i\de^c_a\bar\Psi^q_{\mc A}\mc Z^{\mc A}_b, \\[0.2cm]
\{\bar\Psi_{\mc A\dot r}\Psi^{\mc A}_p,\bar\Psi^q_{\mc B}\mc Z^{\mc B}_a\}_{D.B.}=i\de^q_p\bar\Psi_{\mc A\dot r}\mc Z^{\mc A}_a,\quad
\{R^{\, r}{}_p,\bar\Psi^q_{\mc A}\mc Z^{\mc A}_a\}_{D.B.}=-\frac{i}{2}\de^r_p\bar\Psi^q_{\mc A}\mc Z^{\mc A}_a+i\de^q_p\bar\Psi^r_{\mc A}\mc Z^{\mc A}_a;\\[0.2cm]
\{\bar{\mc Z}^c_{\mc A}\mc Z^{\mc A\dot b},\bar{\mc Z}_{\mc B\,\dot a}\Psi^{\mc B\dot q}\}_{D.B.}=i\de^{\dot b}_{\dot a}\bar{\mc Z}^c_{\mc A}\Psi^{\mc A\dot q}, \quad
\{M_{\dot c}{}^{\dot b},\bar{\mc Z}_{\mc A\,\dot a}\Psi^{\mc A\dot q}\}_{D.B.}=-\frac{i}{2}\de^{\dot b}_{\dot c}\bar{\mc Z}_{\mc A\,\dot a}\Psi^{\mc A\dot q}+i\de^{\dot b}_{\dot a}\bar{\mc Z}_{\mc A\dot c}\Psi^{\mc A\dot q},\\[0.2cm]
\{\bar\Psi_{\mc A\dot r}\Psi^{\mc A}_p,\bar{\mc Z}_{\mc B\,\dot a}\Psi^{\mc B\dot q}\}_{D.B.}=i\de_{\dot r}^{\dot q}\bar{\mc Z}_{\mc A\,\dot a}\Psi^{\mc A}_p,\quad
\{S_{\dot r}{}^{\dot p},\bar{\mc Z}_{\mc A\,\dot a}\Psi^{\mc A\dot q}\}_{D.B.}=-\frac{i}{2}\de^{\dot p}_{\dot r}\bar{\mc Z}_{\mc A\,\dot a}\Psi^{\mc A\dot q}+i\de_{\dot r}^{\dot q}\bar{\mc Z}_{\mc A\,\dot a}\Psi^{\mc A\dot p}; \\[0.2cm]
\{\bar{\mc Z}_{\mc A\dot c}\mc Z^{\mc A}_b,\bar\Psi_{\mc B\dot q}\mc Z^{\mc B\,\dot a}\}_{D.B.}=-i\de_{\dot c}^{\dot a}\bar\Psi_{\mc A\dot q}\mc Z^{\mc A}_b,\quad
\{M_{\dot c}{}^{\dot b},\bar\Psi_{\mc A\dot q}\mc Z^{\mc A\,\dot a}\}_{D.B.}=\frac{i}{2}\de^{\dot b}_{\dot c}\bar\Psi_{\mc A\dot q}\mc Z^{\mc A\dot a}-i\de_{\dot c}^{\dot a}\bar\Psi_{\mc A\dot q}\mc Z^{\mc A\dot b}, \\[0.2cm]
\{\bar\Psi^r_{\mc A}\Psi^{\mc A\dot p},\bar\Psi_{\mc B\dot q}\mc Z^{\mc B\,\dot a}\}_{D.B.}=-i\de^{\dot p}_{\dot q}\bar\Psi^r_{\mc A}\mc Z^{\mc A\,\dot a},\quad
\{S_{\dot r}{}^{\dot p},\bar\Psi_{\mc A\dot q}\mc Z^{\mc A\,\dot a}\}_{D.B.}=\frac{i}{2}\de^{\dot p}_{\dot r}\bar\Psi_{\mc A\dot q}\mc Z^{\mc A\,\dot a}-i\de^{\dot p}_{\dot q}\bar\Psi_{\mc A\dot r}\mc Z^{\mc A\,\dot a}.
\end{array}
\eeq
D.B. relations involving fermionic second-class constraints are
\beq
\begin{array}{c}
\{\bar{\mc Z}^c_{\mc A}\mc Z^{\mc A\dot b},\bar{\mc Z}_{\mc B\dot a}\Psi^{\mc B}_q\}_{D.B.}=i\de_{\dot a}^{\dot b}\bar{\mc Z}^c_{\mc A}\Psi^{\mc A}_q,\quad
\{M_{\dot c}{}^{\dot b},\bar{\mc Z}_{\mc A\dot a}\Psi^{\mc A}_q\}_{D.B.}=-\frac{i}{2}\de^{\dot b}_{\dot c}\bar{\mc Z}_{\mc A\dot a}\Psi^{\mc A}_q+i\de_{\dot a}^{\dot b}\bar{\mc Z}_{\mc A\dot c}\Psi^{\mc A}_q,\\[0.2cm]
\{\bar\Psi^r_{\mc A}\Psi^{\mc A\dot p},\bar{\mc Z}_{\mc B\dot a}\Psi^{\mc B}_q\}_{D.B.}=-i\de^r_q\bar{\mc Z}_{\mc A\dot a}\Psi^{\mc A\dot p},\quad
\{R^{\, r}{}_p,\bar{\mc Z}_{\mc A\dot a}\Psi^{\mc A}_q\}_{D.B.}=\frac{i}{2}\de^r_p\bar{\mc Z}_{\mc A\dot a}\Psi^{\mc A}_q-i\de^r_q\bar{\mc Z}_{\mc A\dot a}\Psi^{\mc A}_p; \\[0.2cm]
\{\bar{\mc Z}_{\mc A\dot c}\mc Z^{\mc A}_b,\bar\Psi^q_{\mc B}\mc Z^{\mc B\,\dot a}\}_{D.B.}=-i\de_{\dot c}^{\dot a}\bar\Psi^q_{\mc A}\mc Z^{\mc A}_b,\quad
\{M_{\dot c}{}^{\dot b},\bar\Psi^q_{\mc A}\mc Z^{\mc A\,\dot a}\}_{D.B.}=\frac{i}{2}\de^{\dot b}_{\dot c}\bar\Psi^q_{\mc A}\mc Z^{\mc A\,\dot a}-i\de_{\dot c}^{\dot a}\bar\Psi^q_{\mc A}\mc Z^{\mc A\dot b},\\[0.2cm]
\{\bar\Psi_{\mc A\dot r}\Psi^{\mc A}_p,\bar\Psi^q_{\mc B}\mc Z^{\mc B\,\dot a}\}_{D.B.}=i\de^q_p\bar\Psi_{\mc A\dot r}\mc Z^{\mc A\,\dot a},\quad
\{R^{\, r}{}_p,\bar\Psi^q_{\mc A}\mc Z^{\mc A\,\dot a}\}_{D.B.}=-\frac{i}{2}\de^r_p\bar\Psi^q_{\mc A}\mc Z^{\mc A\,\dot a}+i\de^q_p\bar\Psi^r_{\mc A}\mc Z^{\mc A\,\dot a};\\[0.2cm]
\{\bar{\mc Z}_{\mc A\dot c}\mc Z^{\mc A}_b,\bar{\mc Z}^a_{\mc B}\Psi^{\mc B\dot q}\}_{D.B.}=i\de_b^a\bar{\mc Z}_{\mc A\dot c}\Psi^{\mc A\dot q},\quad
\{L^c{}_b,\bar{\mc Z}^a_{\mc A}\Psi^{\mc A\dot q}\}_{D.B.}=-\frac{i}{2}\de^c_b\bar{\mc Z}^a_{\mc A}\Psi^{\mc A\dot q}+i\de_b^a\bar{\mc Z}^c_{\mc A}\Psi^{\mc A\dot q}, \\[0.2cm]
\{\bar\Psi_{\mc A\dot r}\Psi^{\mc A}_p,\bar{\mc Z}^a_{\mc B}\Psi^{\mc B\dot q}\}_{D.B.}=i\de_{\dot r}^{\dot q}\bar{\mc Z}^a_{\mc A}\Psi^{\mc A}_p,\quad
\{S_{\dot r}{}^{\dot p},\bar{\mc Z}^a_{\mc A}\Psi^{\mc A\dot q}\}_{D.B.}=-\frac{i}{2}\de^{\dot p}_{\dot r}\bar{\mc Z}^a_{\mc A}\Psi^{\mc A\dot q}+i\de_{\dot r}^{\dot q}\bar{\mc Z}^a_{\mc A}\Psi^{\mc A\dot p}; \\[0.2cm]
\{\bar{\mc Z}^c_{\mc A}\mc Z^{\mc A\dot b},\bar\Psi_{\mc B\dot q}\mc Z^{\mc B}_a\}_{D.B.}=-i\de^c_a\bar\Psi_{\mc A\dot q}\mc Z^{\mc A\dot b},\quad
\{L^c{}_b,\bar\Psi_{\mc A\dot q}\mc Z^{\mc A}_a\}_{D.B.}=\frac{i}{2}\de^c_b\bar\Psi_{\mc A\dot q}\mc Z^{\mc A}_a-i\de^c_a\bar\Psi_{\mc A\dot q}\mc Z^{\mc A}_b, \\[0.2cm]
\{\bar\Psi^r_{\mc A}\Psi^{\mc A\dot p},\bar\Psi_{\mc B\dot q}\mc Z^{\mc B}_a\}_{D.B.}=-i\de^{\dot p}_{\dot q}\bar\Psi^r_{\mc A}\mc Z^{\mc A}_a,\quad
\{S_{\dot r}{}^{\dot p},\bar\Psi_{\mc A\dot q}\mc Z^{\mc A}_a\}_{D.B.}=\frac{i}{2}\de^{\dot p}_{\dot r}\bar\Psi_{\mc A\dot q}\mc Z^{\mc A}_a-i\de^{\dot p}_{\dot q}\bar\Psi_{\mc A\dot r}\mc Z^{\mc A}_a.
\end{array}
\eeq

\end{document}